\documentclass[%
pre,
 reprint,
 amsmath,amssymb,
 aps,
]{revtex4-2}

\usepackage[normalem]{ulem}
\usepackage{soul}
\usepackage{amsmath}
\usepackage{amssymb}
\usepackage{color}
\usepackage{graphicx}%
\usepackage{dcolumn}%
\usepackage{bm}%
\usepackage{verbatim}

\begin{document}

\title{Multi-scale semi-Lagrangian lattice Boltzmann method }%

\author{N. G. Kallikounis}
\affiliation
{Department of Mechanical and Process Engineering, ETH Zurich, 8092 Zurich, Switzerland}
\author{B. Dorschner}
\affiliation
{Department of Mechanical and Process Engineering, ETH Zurich, 8092 Zurich, Switzerland}
\author{I. V. Karlin}\thanks{Corresponding author}
 \email{ikarlin@ethz.ch}
\affiliation
{Department of Mechanical and Process Engineering, ETH Zurich, 8092 Zurich, Switzerland}

\date{\today}%

\begin{abstract}

We present a multi-scale lattice Boltzmann scheme, which adaptively refines particles' velocity space. Different velocity sets, i.e., higher- and lower-order lattices, are consistently and efficiently coupled, allowing us to use the higher-order lattice only when and where needed. This includes regions of either high Mach number or high Knudsen number.
The coupling procedure of different lattices consists of either projection of the moments of the higher-order lattice onto the lower-order lattice or lifting 
of the lower-order lattice to the higher-order velocity space. Both lifting and projection are local operations, which enable a flexible adaptive velocity set. The proposed scheme can be formulated both in a static and an optimal, co-moving reference frame, in the spirit of the recently introduced Particles on Demand method. The multi-scale scheme is first validated through a convected athermal vortex and also studied in a jet flow setup. The performance of the proposed scheme is further investigated through the shock structure problem and a high Knudsen Couette flow, typical examples of highly non-equilibrium flows in which the order of the velocity set plays a decisive role. The results demonstrate that the proposed multi-scale scheme can operate accurately, with flexibility in terms of the underlying models and with reduced computational requirements.
\end{abstract}

\maketitle

\section{Introduction}

With its roots in kinetic theory, the lattice Boltzmann method (LBM) describes the evolution of fluid flow via the propagation and collision of discretized particle distribution functions (populations) $f_i(\bm x, t)$, which 
are associated with a set of discrete velocities and constructed to recover the macroscopic Navier-Stokes equations (NSE) in the hydrodynamic limit.
LBM has matured to a competitive alternative to conventional numerical solvers, with a vast range of applications including compressible flows \cite{Frapolli2015}, complex moving geometries \cite{Dorschner_ComplexGeo}, multiphase flows \cite{Multi_Mazloomi,Multi_Wohrwag} and rarefied gas dynamics \cite{shan2006}, to mention a few.

While LBM has indeed conquered a large range of fluid dynamics, most popular LBM models use so-called standard lattices such as the D2Q9 or the D3Q27 in two or three dimensions ($D=2,3$) with Q=9 and Q=27 discrete velocities, respectively. 
While this is mainly due to their simplicity and efficiency, the limited number of speeds puts severe restrictions on their range of validity. 
On the other hand, a systematic increase of the number of velocities to so-called high-order or multispeed lattices has been shown to extend the range of validity significantly. High-order lattices can be constructed systematically either by discretizing the Boltzmann equation on the roots of the Hermite polynomials \cite{Shan1998,shan2006} or by entropy considerations, yielding a set of so-called admissible lattices \cite{Chikatamarla2006,Chikatamarla2009}. Note that the roots of the Hermite polynomials are irrational numbers and thus require off-lattice propagation schemes such as the semi-Lagrangian LBM \cite{di2018simulation,SemiLag2017}, whereas admissible lattices as in \cite{Chikatamarla2006,Chikatamarla2009} remain on-lattice with integer-valued velocities. The increase of accuracy of such  lattices can be exploited in many applications such high-speed flows, non-equilibrium gas flows or relativistic fluids \cite{Frapolli2015,AnsumaliPlateSlip,Meng_RarefiedAccuracy,Mendoza2010} to name a few.

In this paper, we will restrict our attention to compressible as well as non-equilibrium flows but the proposed concepts are generic and can be used for all 
multi-scale applications using high-order lattices.
In particular, it is well known that the lack of Galilean invariance and insufficient isotropy of standard lattices, limits classical LB models to isothermal, low-Mach number flows \cite{Qian_Orszag_1993,Qian1998} and the extension of LBM to high-speed compressible flows is still an active area of investigation. For instance, so-called augmented LB models have been developed in \cite{Prasianakis2007, Saadat2019} to mitigate these shortcomings by
introducing non-local corrections into the kinetic equations to eliminate the error terms in the momentum and energy equation, which arise due to the constraints of the standard lattices. Promising results have been shown in recent contributions, featuring both variable Prandtl number and adiabatic exponent \cite{Saadat2019}. Moreover, the so-called Particles on Demand (PonD) method has recently been proposed in \cite{Pond,reyhanian2020thermokinetic}, which eliminates the Galilean invariance errors of the standard lattices from the outset by representing the populations in a co-moving reference frame. Note that while both of these approaches get by with minimal velocity sets, another alternative to lift the aforementioned constraints is the use of multispeed lattices. There, the increase of the number of speeds moderates the lattice constraints and pertinent moments to recover the full Navier-Stokes-Fourier (NSF) equations in the hydrodynamic limit can be represented by the lattice \cite{Alexander1993,Kataoka2004,Li2007,Watari2007}. It must be noted however that while 
multispeed lattices can extend the range of velocity significantly, 
the associated temperature range typically decreases with an increase of the number of particle's velocities. Recently, an entropic LBM realization of multispeed lattices has demonstrated promising results for both trans- and supersonic 
flows\cite{Frapolli2015}.

High-order lattices can also be used to increase accuracy in non-equilibrium flows.
The degree of non-equilibrium or rarefaction is usually quantified by the Knudsen number, which is defined as the ratio of the molecular mean free path and a characteristic length. 
It has been shown both analytically and numerically \cite{kim2008slip,SlipPlateMeng} that by increasing the number of discrete velocities (order of Gauss-Hermite quadrature), LB models can capture non-equilibrium effects of wall-bounded flows beyond the NSF level.

With the examples from above in mind, it needs to be mentioned that while high-order lattices can provide a more accurate description of the flow, they come at high computational costs, which can make these models prohibitive
for flows with realistic complexity in three dimensions. Fortunately, for most practical applications, the regions requiring high-order velocity sets are typically confined to a small sub-region of the entire computational domain. 
Hence, significant computational resources can be saved by using a multi-scale description, which uses higher-order lattices only when and where needed.
In that spirit, a variety of different multi-scale frameworks, coupling different methods have been proposed in the literature \cite{MultiscaleBook}. For example, 
a multi-scale model, coupling LBM for continuum regions to direct simulation Monte Carlo (DSMC) for the high Knudsen number regions, was proposed in \cite{Succi_DSMCLBM,Succi_DSMCLBM2} for steady-state simulations. Furthermore, so-called discrete velocity models (DVM) have been shown to be successful in simulating rarefied gases \cite{broadwell_1964,DVM_MIEUSSENS,titarev_2012} and multi-scale schemes with adaptively refined phase space meshes have been proposed 
to reduce both computational time and memory \cite{AdaptiveVel1,AdaptiveVel2}.
DVMs have also been coupled with the more efficient LBM, which was used in continuum flow zones whereas the DVM was restricted to the rarefied regions only \cite{ARISTOV2020,DVMIliyn}. Finally, in the realm of LBM, a finite difference LB scheme for rarefied gas dynamics was proposed in \cite{multiscale}, where different lattices are coupled in a static manner by a non-local extrapolation procedure.  
While interesting, this approach is limited to static phase space refinement and suffers from severe stability issues due to a non-local \emph{ad hoc} coupling procedure of different lattices based on extrapolation procedures. 

In this work, we propose a multi-scale LBM scheme, which alleviates these issues and allows for adaptive phase space refinement using a consistent and local coupling procedure. For maximum lattice flexibility, we use a semi-Lagrangian advection procedure to naturally decouple the velocity space from the physical space \cite{Pond, di2018simulation,Kramer_SemiLagrangian2020,InterBasedLB_Shu,InterpoBasedLB_cHENG_HUNG,Bardow_2006}. Higher-order lattices are thus only used when and where necessary, while  preserving second-order accuracy. We further shed light on the nature of multi-scale problems and the range of validity of coupling procedures. The proposed scheme is then validated on examples in both high-Mach and high-Knudsen number flows but the scheme can be beneficial whenever large lattices are needed in a confined region. For illustration of the coupling procedure, we use a dual population, multispeed LBM model with variable Prandtl number and adiabatic exponent as proposed in \cite{Frapolli2015} for high-Mach regions.
Further, while high-Mach number flows are intrinsically captured in PonD through an adaptive reference frame, we extend PonD's range of validity to non-equilibrium flows using multispeed lattices.

The paper is organised as follows. In Sec.\ \ref{Model_Section} the two-population model is presented and it's semi-Lagrangian realisation is explained. Subsequently, in Sec. \ref{Multiscale_Section} the multi-scale coupling scheme is introduced. Numerical results are presented in Sec.\ \ref{results_Section}, with simulations of an athermal convected vortex, a jet flow, the shock structure problem and a high Knudsen Couette flow. Finally, concluding remarks are given in Sec.\ \ref{Conclusion_section}.

\section{Model Description}
\label{Model_Section}

\subsection{Discrete velocities}
\label{sec:DiscreteVel}

Without a loss of generality, we consider discrete speeds in two dimensions, $D=2$, formed by tensor products of roots of Hermite polynomials $c_{i\alpha}$,
\begin{equation}
	\label{eq:ci}
	\bm{c}_i=(c_{ix}, c_{iy}).
\end{equation}
The roots of the lowest three Hermite polynomials are collected in Tab.\ \ref{tab:GaussHermiteVelSets} for the sake of completeness. Following the standard nomenclature, we refer to the corresponding discrete speeds (\ref{eq:ci}) as $D2Q9$, $D2Q16$ and $D2Q25$ models. 
Each model is characterized by the lattice temperature $T_L$ and the weights $W_i$ associated with the vectors (\ref{eq:ci}),
\begin{equation}
	\label{eq:wi}
	W_i=w_{ix}w_{iy},
\end{equation}
where $w_{i\alpha}$ are weights of the Gauss--Hermite quadrature, see Tab.\ \ref{tab:GaussHermiteVelSets}. Note that the lattice temperature is matched at the outset, $T_L=1$, for all the models under consideration.

\begin{table}[h] \centering
	\caption{Lattice temperature $T_L$, roots of Hermite polynomials $c_{i\alpha}$ and weights $w_{i\alpha}$ of the $D=1$ Gauss--Hermite quadrature, and nomenclature.}
	\label{tab:GaussHermiteVelSets}
	\begin{tabular}{l|l|l|l|l}
		Model & $T_L$   &$c_{i\alpha}$      & $w_{i\alpha}$    & $D=2$     \\ 
		      &         &                   &                   &    \\
		$D1Q3 $ & $1$ &  $0,$     & $2/3$   & $D2Q9$          \\ 
		&    &$\pm\sqrt{3}$   & $1/6$       &       \\ 
		&        &                &          &           \\
		$D1Q4$ & $1$  &$\pm \sqrt{3-\sqrt{6}} $     & $(3+\sqrt{6})/12$   &   $D2Q16$         \\
		&   &$\pm \sqrt{3+\sqrt{6}}$     & $(3-\sqrt{6})/12$           &   \\
		&   &                            &                            &    \\
		$D1Q5$ & $1$   &$0$   & $8/15$          & $D2Q25$    \\ 
		&$ $     &$\pm \sqrt{5-\sqrt{10}}$   & $(7+2\sqrt{10})/60$    &          \\ 
		&$ $     &$\pm \sqrt{5+\sqrt{10}}$   & $(7-2\sqrt{10})/60$    &          \\ 
	\end{tabular}
\end{table}

With the discrete speeds (\ref{eq:ci}), the particles' velocities $	\bm{v}_i$ are defined relative to a reference frame,
specified by the frame velocity $\bm{u}_{{\rm ref}}$ and the reference temperature $T_{{\rm ref}}$,
\begin{equation}\label{eq:veli}
	\bm{v}_i=\sqrt{\frac{RT_{\rm ref}}{T_L}}\bm{c}_i+\bm{u}_{\rm ref},
\end{equation}
where $R$ is the gas constant. Two reference frames of interest shall be considered below. The local co-moving reference frame is specified by the local temperature $T=T(\bm{x},t)$ and the local flow velocity $\bm{u}=\bm{u}(\bm{x},t)$. The lattice reference frame is specified by $\bm{u}_{\text{ref}}=\bm{0}$ and $T_{\text{ref}}=T_L/R$.

\subsection{Kinetic equations}
 
For the sake of presentation, we consider a 
	two-population 
	kinetic model for  ideal gas with a variable 
	adiabatic exponent and 
	Prandtl number \cite{Frapolli2015},
\begin{widetext}
\begin{align}
\label{Pond_f_equation}
f_i(\bm{x},t) - f_i(\bm{x}-\bm{v}_i\delta t,t-\delta t) 
=  \omega_1(f_i^{eq}-f_i)+(\omega_1-\omega_2)(f_i^{\ast}-f_i^{eq}),\\
g_i(\bm{x},t) - g_i(\bm{x}-\bm{v}_i\delta t,t-\delta t) 
=  \omega_1(g_i^{eq}-g_i)+(\omega_1-\omega_2)(g_i^{\ast}-g_i^{eq}), 
\label{Pond_g_equation}
\end{align}
\end{widetext}
where $f_i^{eq}$ and $g_i^{eq}$ are local equilibria while  $f_i^{\ast}$ and $g_i^{\ast}$ are quasi-equilibrium populations. Moreover, $\delta t$ is the time step and $\omega_1$ and  $\omega_2$ are relaxation parameters related to the dynamic viscosity and thermal conductivity \cite{Frapolli2015}, respectively. The local conservation laws for the density $\rho$, momentum $\rho\bm{u}$ and the total energy $\rho E$ are,
\begin{align}
    \rho &= \sum_{i=0}^{Q-1}f_i = \sum_{i=0}^{Q-1}f_i^{eq}, \\
    \rho\bm{u} &=  \sum_{i=0}^{Q-1}\bm{v}_if_i = \sum_{i=0}^{Q-1}\bm{v}_if_i^{eq}, \\
     \rho E &= \sum_{i=0}^{Q-1} \frac{{v}_i^2}{2}f_i
     	+  \sum_{i=0}^{Q-1} g_i  =  \sum_{i=0}^{Q-1} \frac{{v}_i^2}{2}f_i^{eq}+  \sum_{i=0}^{Q-1} g_i^{eq}.
\end{align}
We consider ideal gas with the internal energy of the form $U=C_vT$, where $C_v$ is the specific heat at constant volume. The total energy is,
\begin{equation}
	\label{eq:E}
	\rho E=C_v\rho T + \frac{\rho{u}^2}{2}.
\end{equation}
The quasi-equilibrium populations are,
	\begin{align}
	&	f_i^{\ast} =
	f_i^{eq}+W_i\frac{  \overline{\bm{Q}}:(\theta^{3/2} \bm{c}_{i}\otimes\bm{c}_{i}\otimes\bm{c}_{i} -RT\theta^{1/2} {\rm sym}(\bm{c}_{i}\otimes\bm{1}))}{6(RT)^3}, \label{eq:fqeq}\\
	&	g_i^{\ast} =
	g_i^{eq}+W_i \frac{\theta^{1/2}\overline{\bm{q}}\cdot \bm{c}_{i}}{RT}, 
	\label{eq:gqeq}
	\end{align}
where ${\rm sym}(\dots)$ denotes symmetrization and $\theta$ is the reduced temperature,
\begin{equation}
	\label{eq:theta}
	\theta=\frac{RT}{T_L},
\end{equation}
while the non-equilibrium third-order tensor $\overline{\bm{Q}}$ and the heat flux vector $\overline{\bm{q}}$ are,
\begin{align}
	&\overline{\bm{Q}} = \sum_{i=0}^{Q-1} (\bm{v}_{i}-\bm{u})\otimes (\bm{v}_{i}-\bm{u})\otimes (\bm{v}_{i}-\bm{u}) (f_i-f_i^{eq}),\\
	&\overline{\bm{q}} = \sum_{i=0}^{Q-1} (\bm{v}_{i}-\bm{u})(g_i -g_i^{eq}).
\end{align}

When the local flow velocity $\bm{u}(\bm{x},t)$ and the local temperature $T(\bm{x},t)$ are used to gauge particles' velocities (\ref{eq:veli}),
	\begin{equation}\label{eq:comoving}
		\bm{u}_{\text{ref}}=\bm{u}(\bm{x},t),\ T_{\text{ref}}=T(\bm{x},t),
	\end{equation} 
we say that kinetic equations (\ref{Pond_f_equation}) and (\ref{Pond_g_equation}) are formulated in the co-moving reference frame, where
the equilibrium populations depend only on the density and the temperature,
\begin{align}
	\label{feqPond}
	&  f_i^{eq} = \rho W_i, \\
	\label{geqPond}
	&  g_i^{eq} = \left(C_v-\frac{D}{2}R\right)T\rho W_i.
\end{align}

With the co-moving reference frame and for any of the models of sec.\ \ref{sec:DiscreteVel}, the hydrodynamic limit of the kinetic equations (\ref{Pond_f_equation}) and (\ref{Pond_g_equation}) are the standard equations of compressible gas dynamics, with the dynamic viscosity $\mu$, thermal conductivity $\kappa$ and the bulk viscosity $\xi$ related to the relaxation parameters $\omega_1$ and $\omega_2$ as follows,
\begin{align}
	\mu &= 
\left(\frac{1}{\omega_1} - \frac{1}{2}\right)\rho RT \delta t,   \\
	\kappa & = 
C_p\left(\frac{1}{\omega_2} - \frac{1}{2}\right)\rho RT \delta t, \\
	 \xi &= 
\left(\frac{1}{C_v}-\frac{2}{DR}\right) \mu.
\end{align}
Here $C_p$ is the specific heat of ideal gas at constant pressure, $C_p=C_v+R$. The Prandtl number is defined as ${\rm Pr}=C_p \mu / \kappa$, and the adiabatic exponent is $\gamma=C_p / C_v$. In the following, we set $R=1$, without loss of generality.

\subsection{Semi-Lagrangian realization}

\subsubsection{Co-moving reference frame}
\label{PondMethod}

The implementation of the propagation using the co-moving reference frame requires a transformation between non-equal reference frames which we remind for the sake of completeness \cite{Pond}. Specification of a reference frame shall be denoted $\lambda$,
\begin{align}
	\label{eq:reference}
	\lambda&=\{\bm{u},T\},\\
	\label{eq:vi}
	\bm{v}_i^\lambda&=\sqrt{\theta}\bm{c}_i+\bm{u},
\end{align}
where $\theta$ is the reduced reference temperature (\ref{eq:theta}). In a given reference frame $\lambda$, the $Q$ linearly independent moments of the $f$- and of the $g$-populations are defined as,
\begin{align}
	\label{eq:momf}
	M_{mn}^{\lambda}&=\sum_{i=0}^{Q-1} f_i^{\lambda}\left(\sqrt{\theta} c_{ix}+u_x\right)^m \left(\sqrt{\theta} c_{iy}+u_y\right)^n,\\
	\label{eq:momg}
	N_{mn}^{\lambda}&=\sum_{i=0}^{Q-1} g_i^{\lambda}\left(\sqrt{\theta} c_{ix}+u_x\right)^m \left(\sqrt{\theta} c_{iy}+u_y\right)^n,
\end{align}
where $m,n\in\{0,\dots,\sqrt{Q}-1\}$. Equations (\ref{eq:momf}) and (\ref{eq:momg}) establish a linear map of the $Q$-dimensional population vectors $f^\lambda$ and $g^\lambda$ into the moment vectors $M^\lambda$ and $N^\lambda$. Denoting $\mathcal{M}_{\lambda}$ the $Q\times Q$ the matrix of this map, we write (\ref{eq:momf}) and (\ref{eq:momg}) as,
\begin{align}
	\label{eq:momfshort}
	M^{\lambda}&=\mathcal{M}_\lambda f^{\lambda},\\
	\label{eq:momgshort}
	N^{\lambda}&=\mathcal{M}_\lambda g^{\lambda}.
\end{align}
Consider two reference frames,  $\lambda$ and $\lambda'$. Following \cite{Pond}, populations are transformed based on the principle of independence of the moments on the reference frame,

\begin{align}\label{eq:fmommatch}
	M^{\lambda}&=M^{\lambda'},\\
	\label{eq:gmommatch}
	 N^{\lambda}&=N^{\lambda'}.
\end{align}

With (\ref{eq:fmommatch}) and (\ref{eq:gmommatch}), the populations are transformed from the reference frame $\lambda$ to the reference frame $\lambda'$,
\begin{align}
\label{eq:fll}
    f^{\lambda'}&=\mathcal{G}_{\lambda}^{\lambda'}f^\lambda,\\
\label{eq:gll}
    g^{\lambda'}&=\mathcal{G}_{\lambda}^{\lambda'}g^\lambda,
\end{align}
where the transfer matrix $\mathcal{G}_{\lambda}^{\lambda'}$ reads,
\begin{equation}
	\mathcal{G}_{\lambda}^{\lambda'}=\mathcal{M}_{\lambda'}^{-1}\mathcal{M}_{\lambda}.
\end{equation}
Finally, populations are reconstructed at a point $\bm{x}$ and time $t$ using Lagrange interpolation,
\begin{align}
	\label{reconstructionPond}
	\Bar{f}_i^\lambda(\bm{x},t)=\sum_{s=1}^{p} a_s(\bm{x}-\bm{x}_s)\mathcal{G}_{\lambda_s}^{\lambda}f^{\lambda_s}(\bm{x}_s,t),\\
	\label{eq:reconstructionPondg}
		\Bar{g}_i^\lambda(\bm{x},t)=\sum_{s=1}^{p} a_s(\bm{x}-\bm{x}_s)\mathcal{G}_{\lambda_s}^{\lambda}g^{\lambda_s}(\bm{x}_s,t),
\end{align}
where the summation is carried out over the collocation points $\bm{x}_s$ and $a_s$ are interpolation functions. Below, we use the third-order Lagrange polynomials. Reconstruction (\ref{reconstructionPond}) and (\ref{eq:reconstructionPondg}) takes into account the fact that the reference frames $\lambda_s$ at various collocation points $x_s$ may differ from one another. Thus, the corresponding populations $f^{\lambda_s}$ and $g^{\lambda_s}$ are transformed into a target reference frame $\lambda$  using (\ref{eq:fll}) and (\ref{eq:gll}) prior to the interpolation.

Evaluation of the populations at the monitoring point $\bm{x}$  at time $t$ involves the propagation and the collision steps. In the propagation step, semi-Lagrangian advection is performed using the reconstruction (\ref{reconstructionPond}) and (\ref{eq:reconstructionPondg}) at the departure points of characteristic lines,
\begin{align}
\label{advectionPond}
    f_i(\bm{x},t)&=\Bar{f}_i(\bm{x}-\bm{v}_i \delta t,t-\delta t),\\
\label{eq:advectionPondg}
g_i(\bm{x},t)&=\Bar{g}_i(\bm{x}-\bm{v}_i \delta t,t-\delta t).
\end{align}
Here, particle's velocities $\bm{v}_i$ are defined relative to the co-moving local reference frame at $\bm{x}$ and $t$. In order to find the co-moving reference frame, a predictor-corrector process is executed as follows: In the prediction step, the local reference frame $\lambda_0=\{ \bm{u}_0,T_0 \}$, is initialized using the local flow velocity and the local temperature available from the previous time step, $\bm{u}_0=\bm{u}(\bm{x},t-\delta t), {T}_0={T}(\bm{x},t-\delta t) $. The density, momentum and temperature are consequently computed,
\begin{align}
\label{correctorfieldsrho}
    \rho_1 &=\sum_{i=0}^{Q-1} f_i^{\lambda_0} , \\
 \label{correctorfieldsu}   
    \rho_1 \bm{u}_1 &=\sum_{i=0}^{Q-1} \bm{v}_i^{\lambda_0} f_i^{\lambda_0}, \\
     \rho_1 {E}_1 &=\sum_{i=0}^{Q-1} \frac{{(\bm{v}_i^{\lambda_0})}^2}{2} f_i^{\lambda_0} +\sum_{i=0}^{Q-1}g_i^{\lambda_0}.
\end{align}
The computed velocity  (\ref{correctorfieldsrho}) and temperature (\ref{correctorfieldsu})  define the corrector reference frame  $\lambda_1=\{ \bm{u}_1,T_1 \}$ at the monitor point and the propagation step (\ref{advectionPond}) is repeated with the updated reference frame. The predictor-corrector process is iterated until convergence with the limit values, 
$$\rho (\bm{x},t),\bm{u} (\bm{x},t),T (\bm{x},t), f_i^{\lambda(\bm{x},t)} =\lim_{n \to \infty} \rho_n,\bm{u}_n, T_n, f_i^{\lambda_n}, $$
defining the density, velocity, temperature and the pre-collision populations at the monitoring point $\bm{x}$ at time $t$. The predictor-corrector iteration loop ensures that the propagation and the collision steps are performed at the co-moving reference frame, in which the local equilibrium populations (\ref{feqPond}, \ref{geqPond}) are exact.

\subsubsection{Lattice reference frame}

If instead of an adaptive co-moving reference PonD frame, the fixed lattice reference frame $\lambda_L=\{ \bm{0},T_{L}\}$ is used, we arrive at  the semi-Lagrangian LBM \cite{SemiLag2017,Kramer_SemiLagrangian2020,di2018simulation}. In this special case, the lattice equilibrium populations $f_i^{L}$, $g_i^{L}$ are evaluated by transforming the exact equilibrium populations (\ref{feqPond}) and (\ref{geqPond}) to the 
lattice reference frame,
\begin{align}
	\label{eq:feqL}
&    f^L= \mathcal{G}_{\{ \bm{u},T\}}^{\{ \bm{0},T_{L}\}}f^{eq},\\
\label{eq:geqL}
&    g^L = \left(C_v-\frac{D}{2}\right)T f^L.    
\end{align}
Similarly, the quasi-equilibria are transformed to the lattice reference frame,
\begin{align}
	\label{eq:feqL}
&    f^{\ast,L}= \mathcal{G}_{\{ \bm{u},T\}}^{\{ \bm{0},T_{L}\}}f^{\ast},\\
\label{eq:geqL}
&    g^{\ast,L} = \mathcal{G}_{\{ \bm{u},T\}}^{\{ \bm{0},T_{L}\}}g^{\ast}.    
\end{align}

A discussion is in order here. It is well known that the standard $D2Q9$ lattice with the lattice equilibrium does not allow for a compressible model. At the same time, the mere change to the co-moving reference frame readily provides a compressible model with the same number of discrete velocities. In order to understand why the choice of the reference frames matters, we remind that attention should be paid at the higher-order moments that are not included in the transformation. For the $D2Q9$ lattice these are the diagonal elements of the third-order moment and the diagonal elements of the fourth-order moment,
\begin{align}
		Q_{\alpha\alpha\alpha}^{\lambda}&=\sum_{i=0}^{8}\left(v^\lambda_{i\alpha}\right)^3 f_i^\lambda,\\
		R_{\alpha\alpha\alpha\alpha}^{\lambda}&=\sum_{i=0}^{8}\left(v^\lambda_{i\alpha}\right)^4 f_i^\lambda.
\end{align}
While the off-diagonal third-order elements, $Q_{xyy}=M_{12}$ and $Q_{yxx}=M_{21}$, as well as the off-diagonal fourth-order element $R_{xy}=M_{22}$ are the same in both, the co-moving and the lattice reference frame since they are among the moments transformed, the equilibrium values of the non-transformed moments are not equal in both references. Indeed, using the co-moving equilibrium (\ref{feqPond}) we find,
\begin{align}
	\label{eq:QxxxPonD}
	& Q_{\alpha\alpha\alpha}^{eq}=3\rho T u_\alpha+\rho u_\alpha^3,\\
	\label{eq:RxxPond}
	&	R_{\alpha\alpha\alpha\alpha}^{eq}=3\rho T^2 + 6\rho T u_\alpha^2 +\rho u_\alpha^4.
\end{align}
With the diagonal moments (\ref{eq:QxxxPonD}) and (\ref{eq:RxxPond}), the equilibrium third-order moment tensor and the once-contracted equilibrium fourth-order moment tensor become,
\begin{align}
	\label{eq:QPonD}
Q^{eq}_{\alpha\beta\gamma}&=\rho T\left(u_{\alpha}\delta_{\beta\gamma}+u_{\beta}\delta_{\alpha\gamma}+u_{\gamma}\delta_{\alpha\beta}\right)+\rho u_{\alpha}u_{\beta}u_{\gamma},\\
\label{eq:RPonD}
\begin{split}
R^{eq}_{\alpha\beta} &=\rho T((D+2)T+u^2)\delta_{\alpha\beta}\\
&+\rho((D+4) T+u^2) u_{\alpha}u_{\beta}.
\end{split}
\end{align}
The equilibrium Maxwell--Boltzmann relations (\ref{eq:QPonD}) and (\ref{eq:RPonD}) are precisely what is required to recover the compressible flow equations in the thermodynamic limit. On the contrary, the lattice equilibrium (\ref{eq:feqL}) returns, instead of (\ref{eq:QxxxPonD}) and (\ref{eq:RxxPond}), 
\begin{align}
	\label{eq:QxxxL}
	& Q_{\alpha\alpha\alpha}^{L}=3\rho T_L  u_\alpha,\\
		\label{eq:RxxL}
	&	R_{\alpha\alpha\alpha\alpha}^{L}=3\rho (T_L)^2 + 3\rho T_L u_\alpha^2.
\end{align}

Finally, if the temperature is fixed to the lattice temperature $T_L$, the athermal model is recovered with
\begin{align}
	\label{eq:feqL0}
	&    f^L= \mathcal{G}_{\{ \bm{u},T_L\}}^{\{ \bm{0},T_{L}\}}f^{eq}. 
\end{align}

\section{Multi-scale coupling scheme}
\label{Multiscale_Section}

\subsection{The overlap and switching}

\begin{figure}
	\centering

	\includegraphics[width=0.5\textwidth]{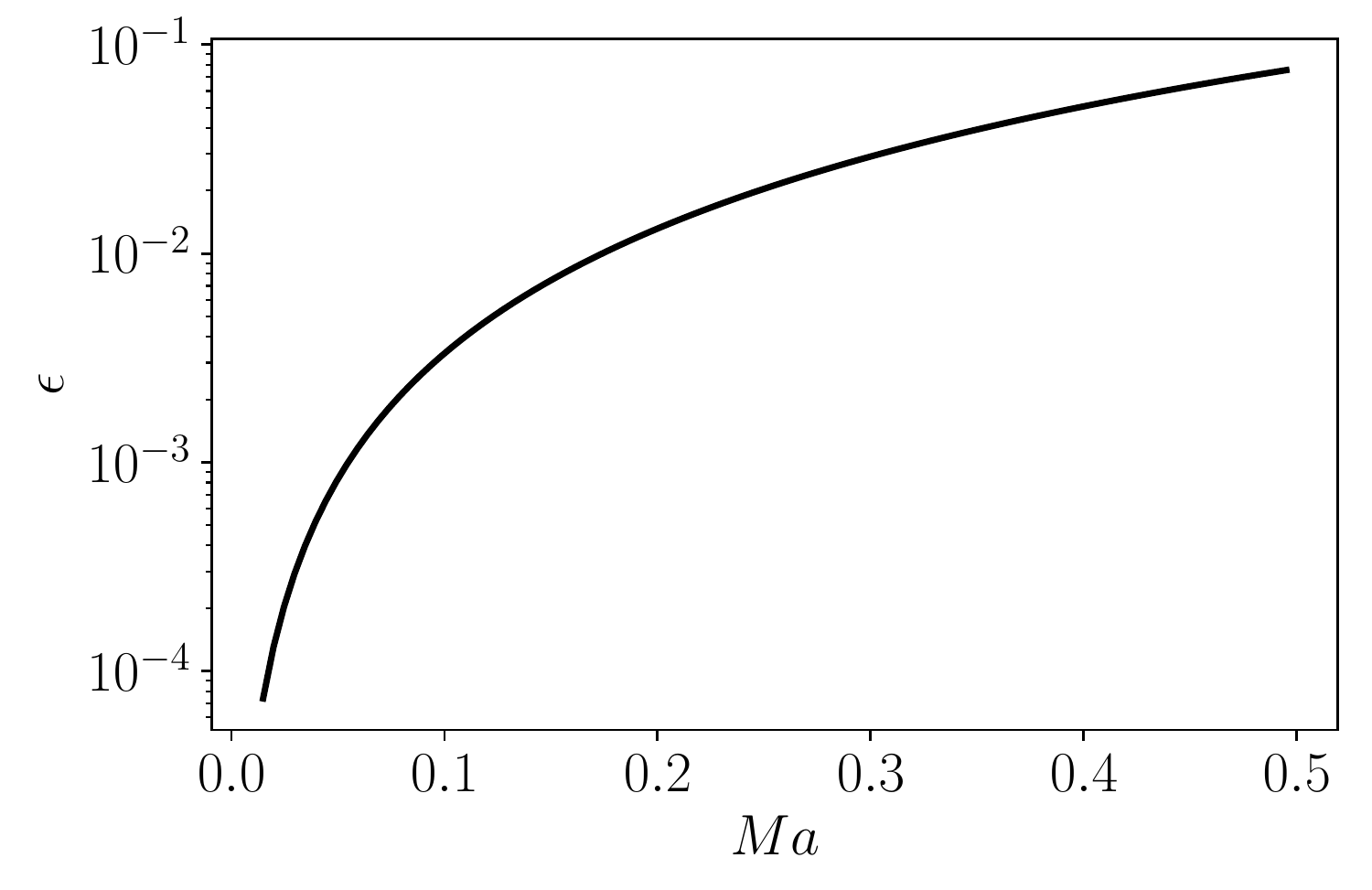} 
	
	\caption{ Deviation of the $xxx$ component of the $D2Q9$ equilibrium third order tensor $Q_{xxx}^{L}$ (\ref{eq:QxxxL}) from it's Maxwell--Boltzmann counterpart $Q_{xxx}^{\text{eq}}$ (\ref{eq:QPonD}) as a function of Mach number ${\rm Ma}=u/c_s$. $\epsilon=(Q_{xxx}^{L}-Q_{xxx}^{\text{eq}})/Q_{xxx}^{\text{eq}}$. The deviation vanishes for the case of the $D2Q25$ lattice.}
	\label{fig:D2Q9error}
\end{figure}

Each discrete velocity set can be characterized by its domain of validity. 
Going by an example, the standard $D2Q9$ lattice provides a reliable simulation of a nearly-incompressible flow as long as the magnitude of the flow velocity stays below $u\lesssim 0.1$ in lattice units. This estimate is based on the deviation of the diagonal element of the equilibrium third-order moment (\ref{eq:QxxxL}) from the Maxwell--Boltzmann value (\ref{eq:QPonD}), see Fig.\ \ref{fig:D2Q9error}. The next-order velocity set in the Hermite hierarchy $D2Q25$ of Tab. \ref{tab:GaussHermiteVelSets} provides a larger domain of validity \cite{shan2006}. It is therefore expected that the use of the $D2Q25$ velocity set instead of the $D2Q9$ can improve the accuracy of the simulation in certain cases. 
At the same time, using the higher-order in the entire computational domain is computationally demanding and
wasteful in regions where the accuracy of the lower-order model is sufficient. 
Because the semi-Lagrangian propagation is not bound to a specific lattice and both the higher- and the lower-order model can be realized on the same grid, we seek an adaptive realization where a higher-order model is used only if the flow situation demands more accuracy than expected from the lower-order model while the lower-order model is used then its accuracy suffices.

Let us consider two velocity sets of different order
	 
		\begin{align*}
			\mathcal{V}_q&=\{\bm{v}_{i}^{q}, i=0, \dotsc, q-1\},\\
			\mathcal{V}_Q&=\{\bm{v}_{i}^{Q}, i=0, \dotsc, Q-1\},
		\end{align*} 
	where $q<Q$.

At a monitoring point $\bm{x}$, at time $t$, we can choose between the two corresponding population vectors, $f_q$ and $f_Q$. 
	We further assume that at $(\bm{x},t)$ the validity domains of the higher-order $Q$-model and the lower-order $q$-model overlap and both models are equally applicable. 
	Switching between the models then requires one of the two operations:
	\begin{itemize}
		\item Lifting: The lifting operation switches from the lower-order $q$-model to the higher-order $Q$-model and is required to increase the accuracy if the available state $f_q$ is close to the limit of its validity domain and we need to proceed with the higher-order model. The lifting operation thus requires us to specify a map,
		\begin{align}
		f_q \to f_Q.	\label{eq:lifting1}
		\end{align}
		\item Projection: The projection operation switches from the higher-order $Q$-model to the lower-order $q$-model and is required to maintain the accuracy with lesser velocities if the available state $f_Q$ is within the validity domain and we can to proceed with the lower-order model. The projection operation thus requires us to specify a map,
		\begin{align}
			f_Q \to f_q.	\label{eq:projection1}
		\end{align}
	\end{itemize}

Note that the lifting and projection must occur when both low- and high-order are equally valid and a proper transfer of information between the different models is possible, i.e., the 
validity domains must overlap. 
This is a general feature that the models of any multi-scale method must respect, such that a coupling strategy can be physically devised. We shall now specify the two switching operations separately.

\subsection{Lifting} 
\label{Reconstruction_Multiscale}
We use notation $m_q$ for the $q$-dimensional vector of moments of the low-order model, available at $(\bm{x},t)$.
	With the $q\times q$ matrix $\mathcal{M}_q$, we have
\begin{align}
		m_q=\mathcal{M}_q f_q.
\end{align}
On the other hand, a moment vector of the $Q$-populations $M_Q$ can be considered as an element of a direct sum,

	\begin{align}
		{M}_{Q}&={M}_q \oplus{M}_{Q-q},
	\end{align}

where $M_q\in {\rm Im}(\mathcal{M}_q)$ is a vector from the image of the matrix  $\mathcal{M}_q$ while ${M}_{Q-q}$ is the rest of the higher-order moments.
The lifting operation consists in specifying the individual contributions as,

\begin{align}
	&M_q=m_q,\\
	&M_{Q-q}=M_{Q-q}^{\rm eq}.
\end{align}

This construction can be understood as a decomposition between conserved, "slow" and "fast" components. The information of the low-order population vector are fully retained, both equilibrium and non-equilibrium contributions, and the moments $M_q$ are identified as the "slow" components. On the other hand, the missing $M_{Q-q}$ are considered as "fast" moments, which are strongly enslaved by the dynamics of the slow moments. The final moment vector $M_{q \to Q}$ can then be written as
\begin{align}
	{M}_{q\to Q}={m}_q \oplus {M}_{Q-q}^{\rm eq}.
\end{align}

With the moments $M_{q \rightarrow Q}$ completed, the populations $f_Q$ are found by moment inversion,

\begin{align}
	f_Q={\mathcal{M}_Q}^{-1} M_{q \to Q},\label{eq:liftingF}
\end{align}

where $\mathcal{M}_Q$ is the $Q\times Q$ moments-to-populations transform for the higher-order model.

\subsection{Projection}

In the projection step, a high-order population vector $f_Q$ is mapped to a lower-order $f_q$. Fortunately, the high-order lattice can represent $Q$ linearly independent moments, which contains the subset of the first $q$ moments, which are required to construct the populations $f_q$ of the low-order lattice. Hence, in contrast to the lifting procedure, there is no missing information and all linearly independent moments $m_{Q \rightarrow q}$ are operationally available from $f_Q$,

\begin{align}
	m_{Q \to q}=M_q.
\end{align}

The population vector $f_q$ is obtained by moment inversion,

\begin{equation}
    f_q={\mathcal{M}_q}^{-1} m_{Q \to q}.
\end{equation}
The concepts of the two mappings are generic and apply with either f (density, momentum conserving lattice) or g (energy conserving lattice) populations. Finally, we stress that both lifting and projection are fully local operations, which leads to high efficiency, numerical stability and flexibility for an adaptive velocity refinement.

\subsection{Semi-Lagrangian propagation}

\begin{figure}
 \centering

\includegraphics[width=0.45\textwidth]{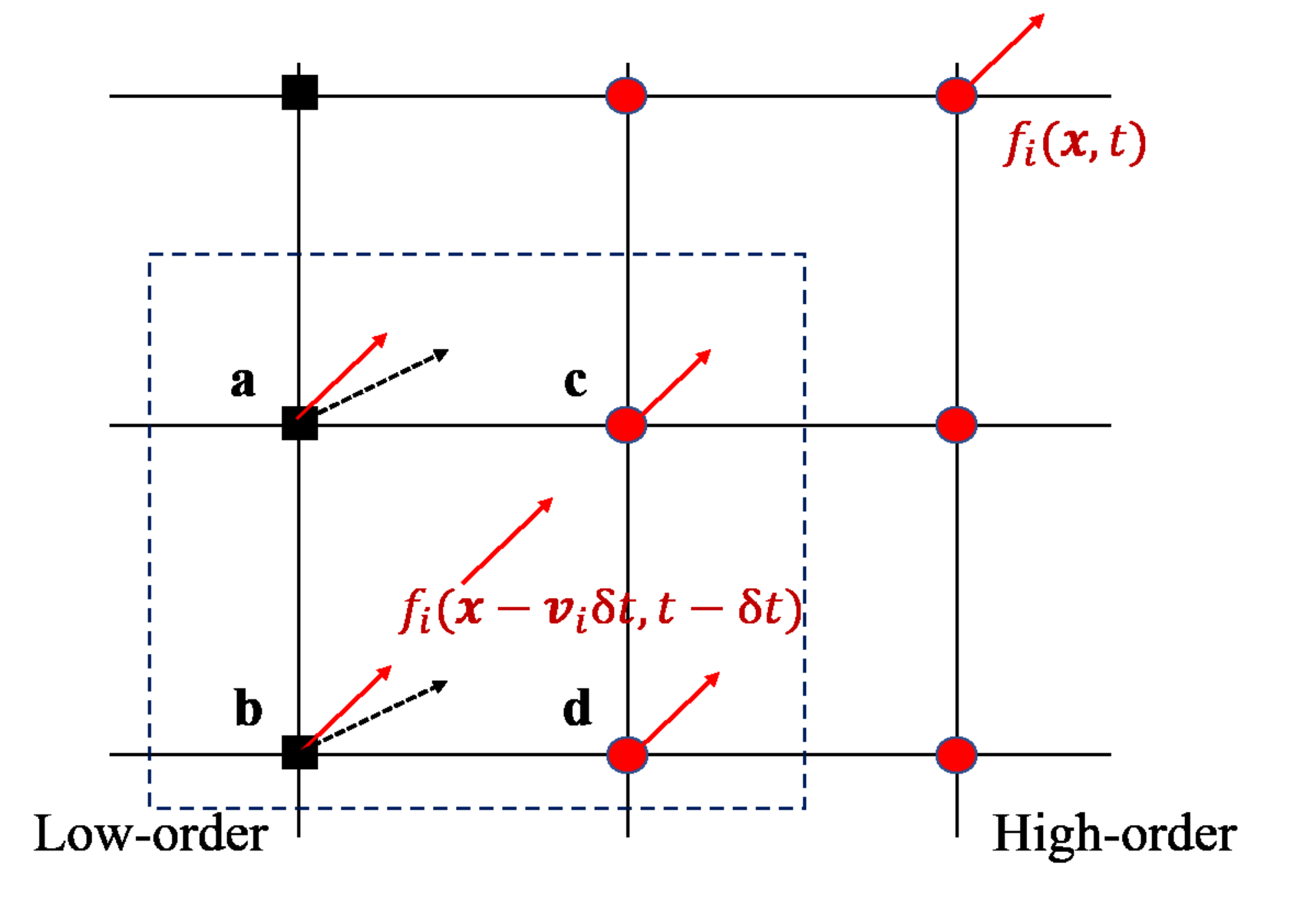} 

    \caption{Schematic of the multi-scale semi-Lagrangian propagation. For simplicity we use four collocating points for the interpolation (inside dotted rectangle) and we consider the case where the lifting operation is active. The departure point $\bm x_{\text{dep},i}=\bm x-\bm v_i\delta{t}$ is surrounded by the collocating points $\bm{a}$ to $\bm{d}$. The black arrows of points $\bm{a}$ and $\bm{b}$ represent the low-order population vector and with the lifting operation the corresponding population in the higher-order velocity set is approximated (red arrows). The semi-Lagrangian step concludes with the interpolation of the populations at the collocating points.}

\label{fig:CouplingScheme}

\end{figure}

Now we discuss the semi-Lagrangian propagation in the multi-scale setting, using the lifting and the projection operators. We consider the population $f_i(\bm{x},t)$, with a discrete velocity $\bm v_i$, belonging to either the lower- or higher-order velocity set. The semi-Lagrangian propagation starts with the calculation of the departure point $\bm x_{\text{dep},i}=\bm x-\bm v_i\delta{t}$, and the identification of the surrounding collocation points, (see, Eq.~\eqref{reconstructionPond}). Here we must take into consideration that the collocation points use in general velocity sets of different order. Thus, the lifting and projection operators are applied, such that the populations involved in the reconstruction are all expressed in the same velocity set. Figure \ref{fig:CouplingScheme} shows an example of the semi-Lagrangian propagation in the multi-scale setting.

The implementation can be summarised in the following steps:
\begin{enumerate}
\item Calculation of departure point, $\bm x_{\text{dep},i}=\bm x-\bm v_i\delta{t}$ and identification of the collocation points.

\item For all collocation points:
\begin{itemize}
    \item {If velocity set order is lower than velocity set order of $(\bm{x},t)$: apply lifting.}
    \item If velocity set order is higher than velocity set order of $(\bm{x},t)$: apply projection.
\end{itemize}

\item Calculation of $f_i(\bm x_{\text{dep},i},t-\delta t)$, using the reconstruction formula Eq.~\eqref{reconstructionPond}.
\end{enumerate}

The first and third step are the same as for the case of a single velocity set throughout the domain, whereas the second step is active only near the interface of different velocity sets. 
The collision step that follows the propagation proceeds as in the case of a uniform velocity set. The modification of the semi-Lagrangian algorithm that enables the multi-scale feature is independent of the underlying model and can be used with single or double populations and with static or co-moving reference frame (PonD). It must be noted that in a co-moving reference frame, the semi-Lagrangian propagation step is performed iteratively in a predictor-correct loop. During the iterations the lifting/projection operations need to be performed once to each point that is close to the interface region (step 2 above). 

\section{Numerical Results and Discussion}
\label{results_Section}

In this section we investigate the performance and accuracy of the multi-scale scheme. The velocity sets that are used in the simulations, generated by Gauss-Hermite quadrature, are described in section \ref{sec:DiscreteVel}. The time step $\delta t$ of the simulations that follow is such that $\text{CFL}= (\text{max}|c_i|\delta t) /\delta x =1.0$, where $c_i$ correspond to the lattice velocities.

\subsection{Multi-scale flows}

Before going into to the details of our validation, it is instructive to remind the nature of multi-scale problems. Multi-scale flows are characterized by large variations of characteristic quantities which most commonly includes but is not limited to spatial scales, time scales, Mach number or Knudsen number. In what follows, we only consider the Mach or Knudsen number, since in those cases we will benefit from using high-order lattices. For large spatio-temporal scales on the other hand, we refer to existing grid-refinement techniques such as \cite{dorschner2016grid}. However, when devising a numerical scheme, which bridges such large variations by coupling different methods for different scales, it is important to realize that consistent coupling is only possible if there is a region, where the validity range of both schemes overlap. 
In our case, this corresponds to regions of the flow, where both lattices can provide an accurate description of the flow field. These regions will eventually become the interface region between both lattices and thus correspond to the phase space refinement criterion. 
Hence, if such overlap regions do not exist in the flow, the use of a multi-scale scheme is inappropriate.

We will first test these ideas for variation in Mach using an athermal convected vortex and an athermal jet flow using a fixed reference frame ($\lambda=\{ \bm{0},T_{L}\}$) as examples. The underlying model for these simulation consist of a single population, without quasi-equilibrium relaxation. The motivation of using different velocity sets in the case of a fixed reference frame stems from the errors of the D2Q9 lattice due to non-Galilean invariance as the Mach number grows and the refinement criterion between high and low order velocity sets is based on a Mach number threshold. We proceed with the shock structure problem using a co-moving reference frame (PonD) and the full compressible model (f,g populations with variable Prandtl number and adiabatic exponent), in which the velocity set is dictated by the variation of the Knudsen number.

\subsubsection{Athermal vortex convection}
\label{sec:vortex}

The reference case of an athermal, convected vortex is studied with the proposed multi-scale scheme, using the standard D2Q9 lattice and the D2Q25. The initial conditions of the velocity and density fields are the following,
\begin{align}
    u_x&=U_0-\epsilon \left(\frac{y-y_c}{R_c}\right)\exp\left[-\frac{(x-x_c)^2+(y-y_c)^2}{2R_c^2}\right],\\
    u_y&=\epsilon \left(\frac{x-x_c}{R_c}\right)\exp\left[-\frac{(x-x_c)^2+(y-y_c)^2}{2R_c^2}\right],\\
    \rho&=\rho_0\exp\left[-\frac{{Ma_v}^2}{2}\exp\left(-\frac{r^2}{R_c^2}\right)\right],
\end{align}
where $U_0$ is the advection velocity, $\epsilon$ is the strength of the vortex and $R_c$ is the characteristic radius \cite{Vortex}. The advection and vortex Mach numbers are set to 0.25 and 0.6 respectively.

\begin{figure}
 \centering

\includegraphics[width=0.5\textwidth]{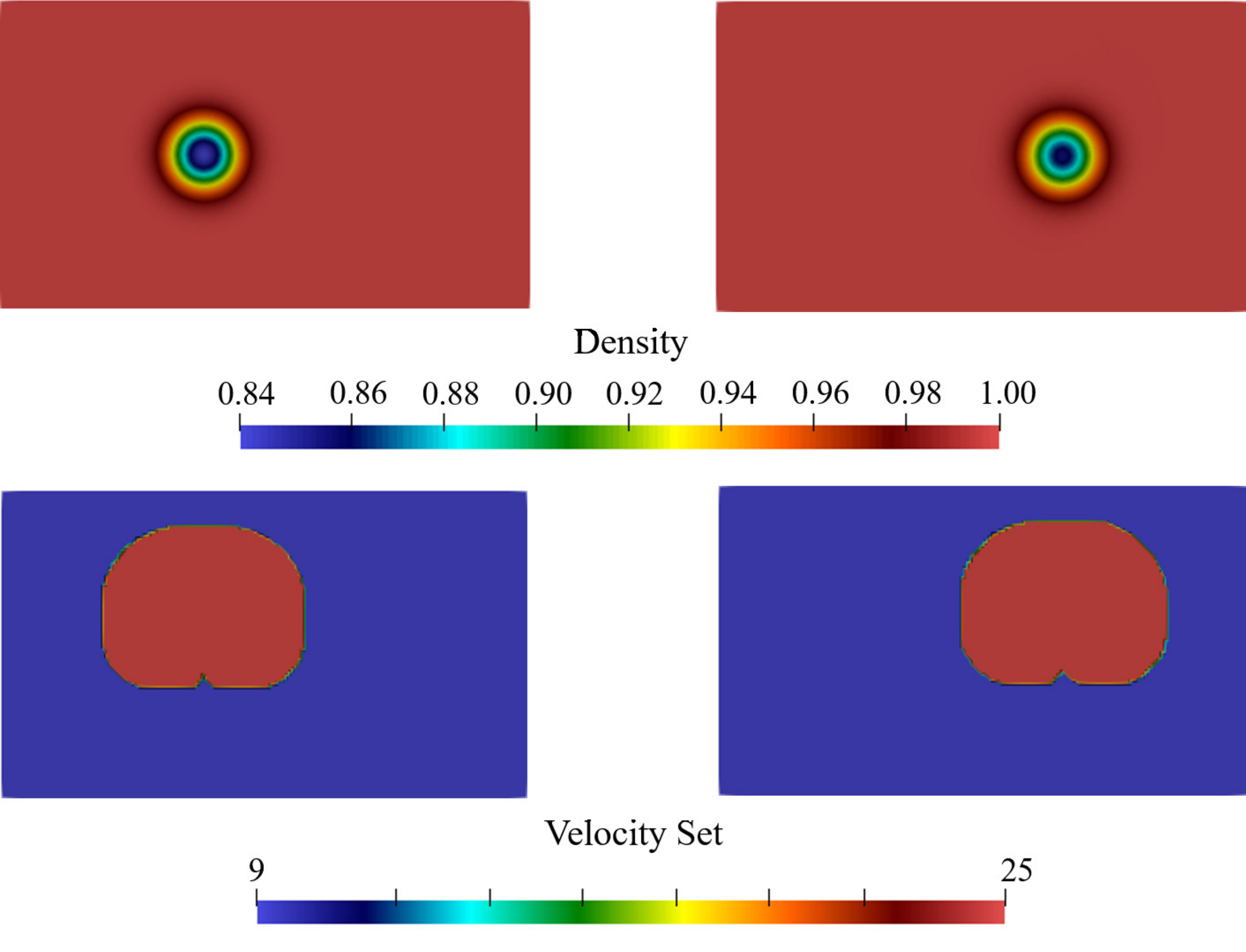} 

\caption{ Density (top) and velocity set (bottom) at two different times ($\Delta t=500$ iterations). The high order D2Q25 follows the vortex as it is advected.}
\label{fig:VortexVelSet}

\end{figure}

For the simulations a $[200 \times 200]$ grid was used with periodic boundary conditions. An estimate for the Mach threshold value that should be used as an appropriate refinement criterion can be obtained from an error analysis of the pertinent equilibrium moments. For athermal, incompressible flows the accuracy of the D2Q9 lattice is limited by the equilibrium third-order moment tensor, $Q_{\alpha \beta \gamma}^{\text{eq}}$. Figure \ref{fig:D2Q9error} shows the deviation of the $Q_{xxx}^{\text{eq}}$ component from it's Maxwell-Boltzmann counterpart $Q_{xxx}^{\text{eq,MB}}=3\rho c_s^2+\rho u^3$ as a function of Mach number, where $c_s$ is the speed of sound. Here a threshold value $Ma_{\text{thr}}=0.3$ was selected, which leads to $3\%$ accuracy. Figure \ref{fig:VortexVelSet} shows the velocity set and the density field at two different times. The high-order velocity set is wrapped around the high speed region of the vortex and follows it during advection.

\begin{figure}
 \centering

\includegraphics[width=0.5\textwidth]{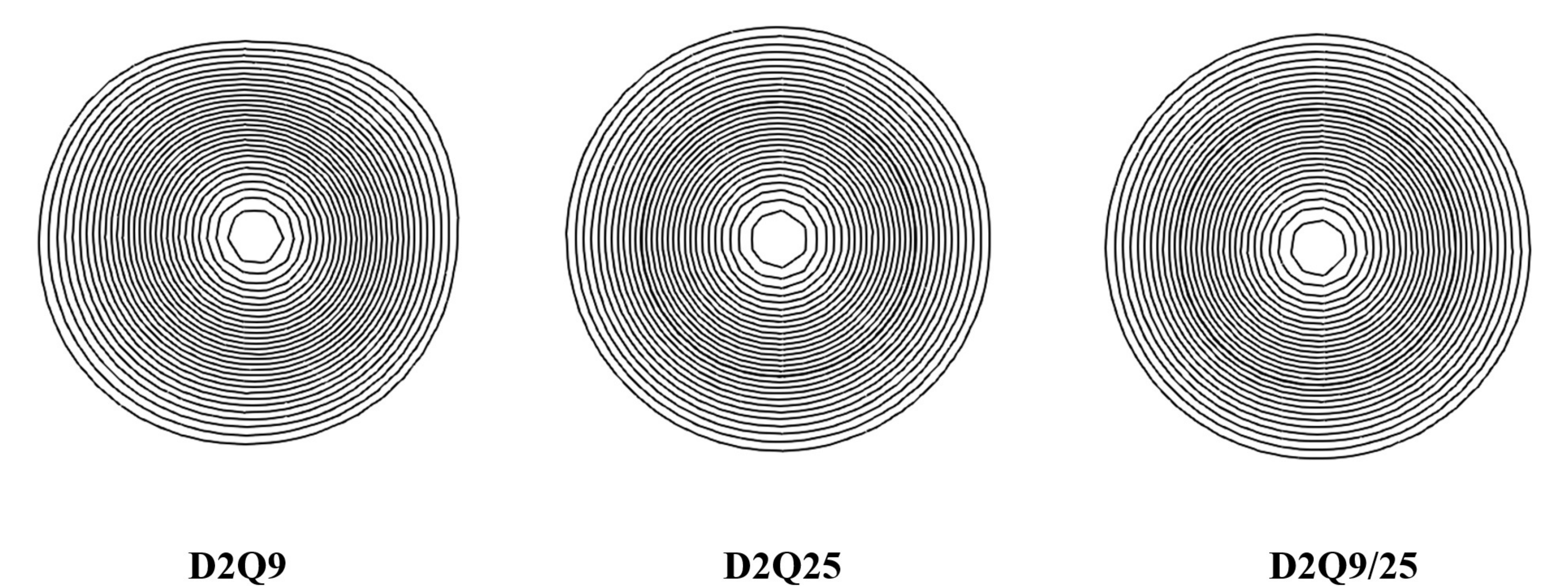} 

\caption{ Density contours of an athermal convected vortex simulation with  D2Q9, D2Q25 and coupled D2Q9/25.}
\label{fig:VortexIso}

\end{figure}

\begin{figure}
 \centering

\includegraphics[width=0.5\textwidth]{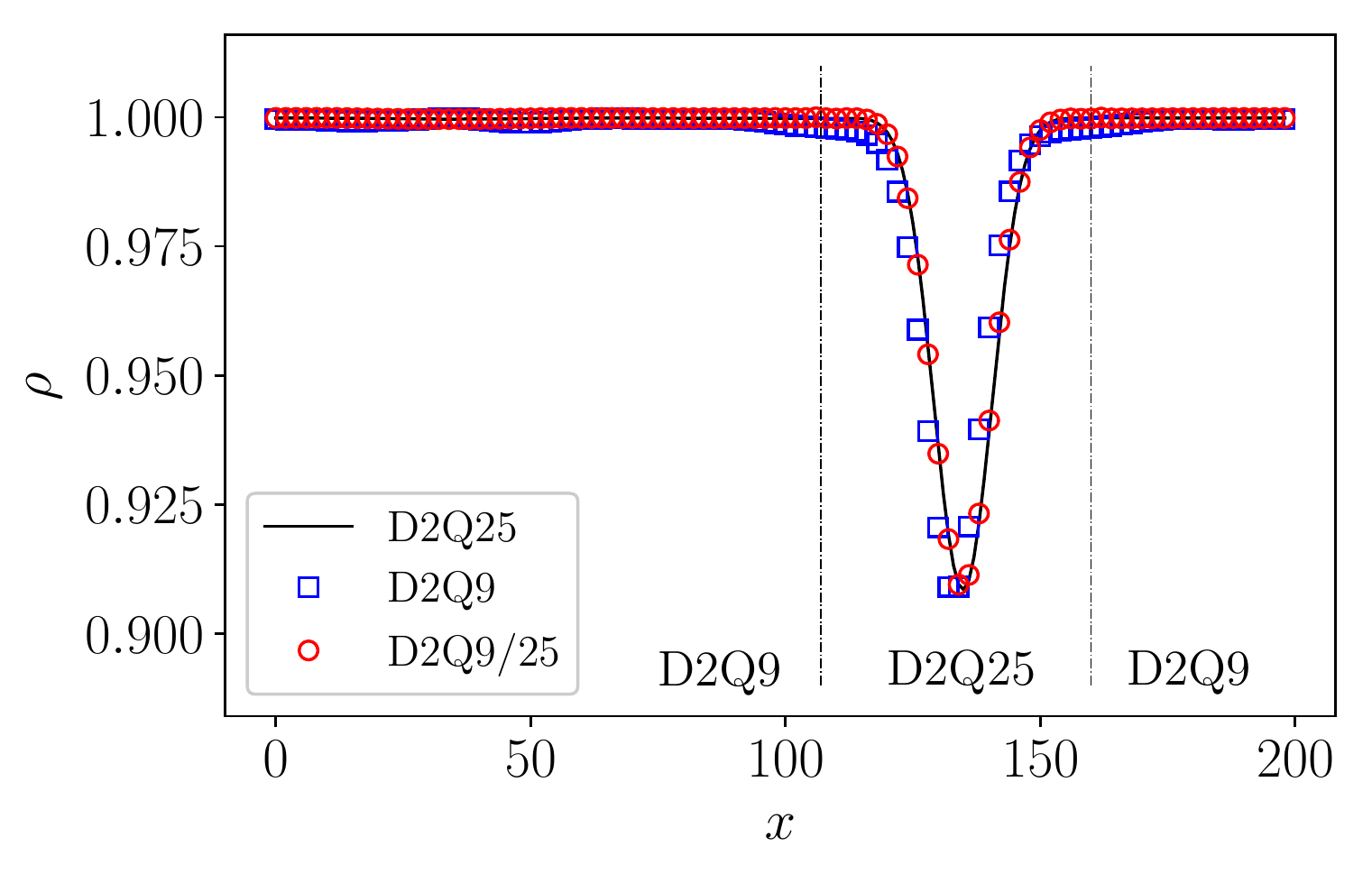} 

\caption{Density of an athermal convected vortex simulation with  D2Q9, D2Q25 and coupled D2Q9/25. The vertical dotted lines on the plot indicate the high-order lattice for the case of the coupled scheme.}
\label{fig:VortexComp}

\end{figure}

The behaviour of the coupling scheme is compared with similar simulations in which a single velocity set is used throughout the domain. In figure \ref{fig:VortexIso}, density contours are shown for the cases of D2Q9, D2Q25 and the hybrid D2Q9/25 velocity set after 500 time steps. The insufficient isotropy of the D2Q9 velocity set manifests in the distortion of the vortex, in contrast with the cases of D2Q25 and the coupled case. In figure \ref{fig:VortexComp} the density profiles are plotted along the center of the vortex, indicating an almost identical behaviour for the D2Q25 and the D2Q9/25 hybrid framework, and deviations for the case of the standard D2Q9 velocity set.

\subsubsection{Athermal jet flow}

The next case to be analysed through the multi-scale scheme is an athermal jet flow. The inflow velocity profile consists of a base low-speed flow with Mach number equal to $Ma_L=0.15$ and a high-speed jet region superimposed symmetrically in the centre of the domain with diameter $D_{\text{jet}}$ and Mach number $Ma_H=0.6$. 
At the inlet a zero pressure gradient and fixed velocity are imposed, whereas a non-reflecting boundary condition is prescribed at the outlet. Free-stream boundary conditions are applied at the top and bottom planes. The simulation is carried out on a $[750 \times 500]$ grid with a jet diameter $D_{\text{jet}}=25$. The viscosity is adjusted such that the jet Reynolds number is $Re={D_{\text{jet}}U_{\text{jet}}}/{\nu}= 1000$.

\begin{figure}

\includegraphics[width=0.45\textwidth]{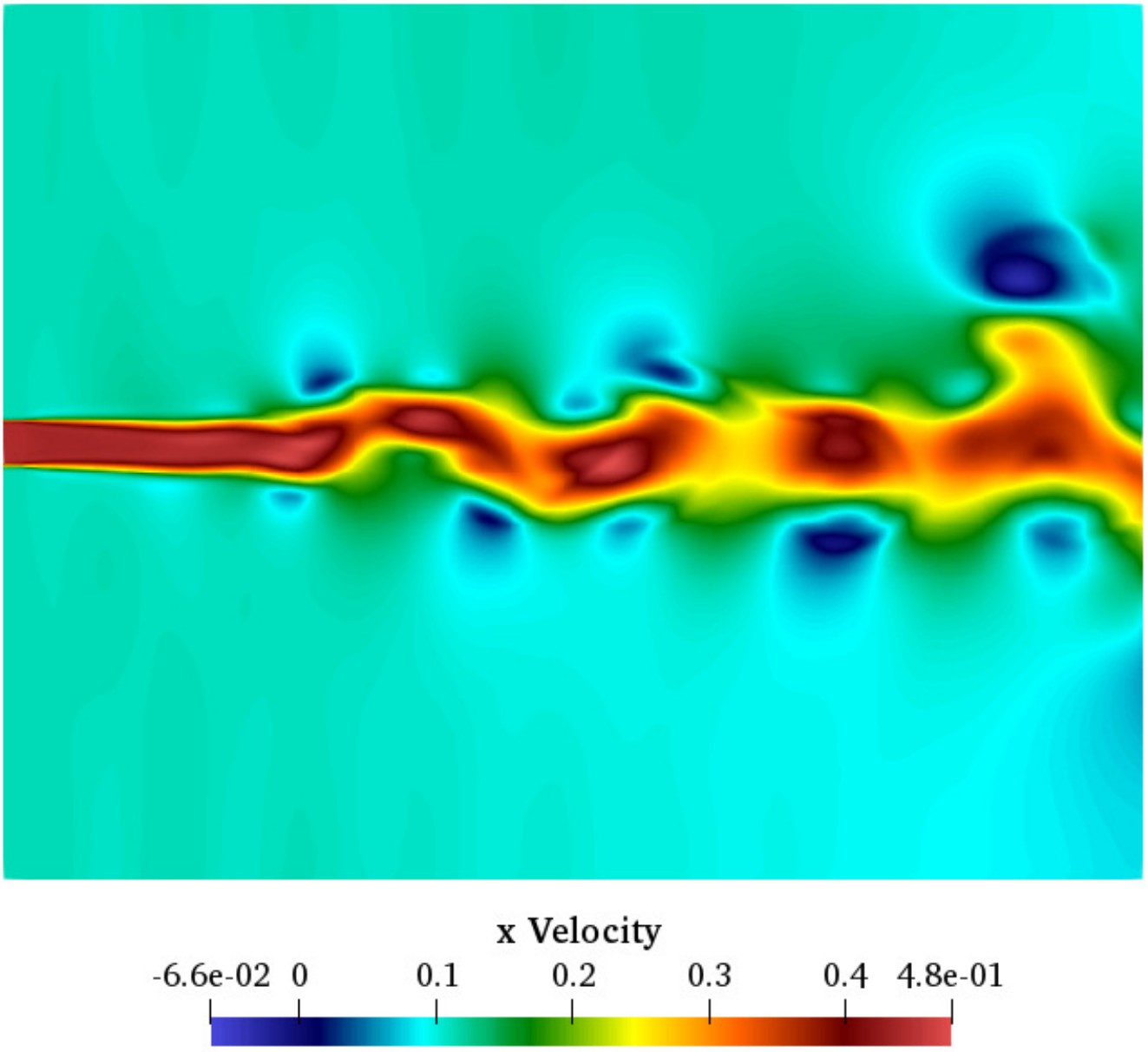} 
\includegraphics[width=0.45\textwidth]{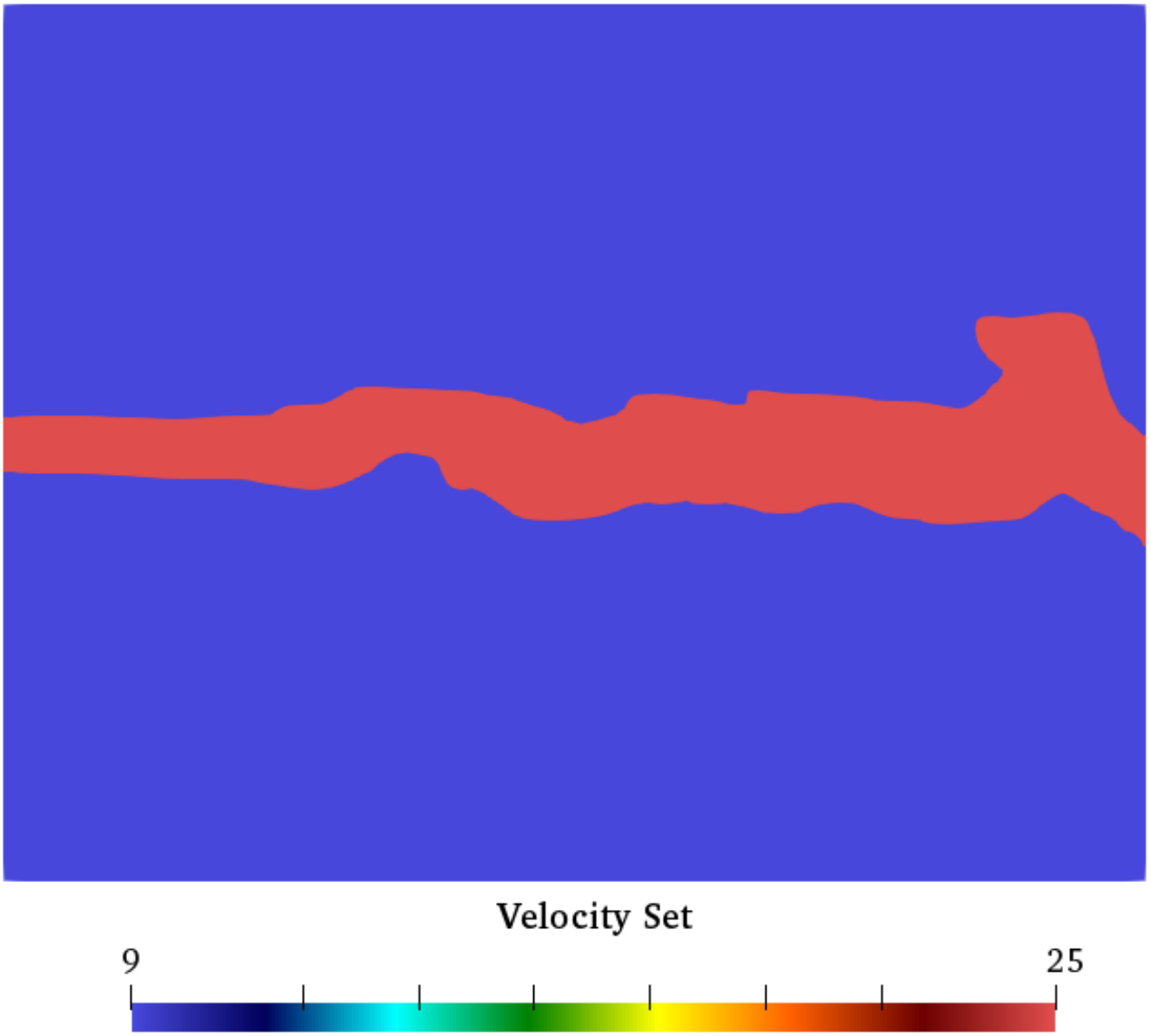}

\caption{Simulation of athermal jet flow with coupled D2Q9/D2Q25. (Top) instantaneous snapshot of the x-velocity field; (Bottom) velocity set throughout the domain at the same time.  }
\label{fig:JetFlow_Inst}
\end{figure}

\begin{figure}

\includegraphics[width=0.45\textwidth]{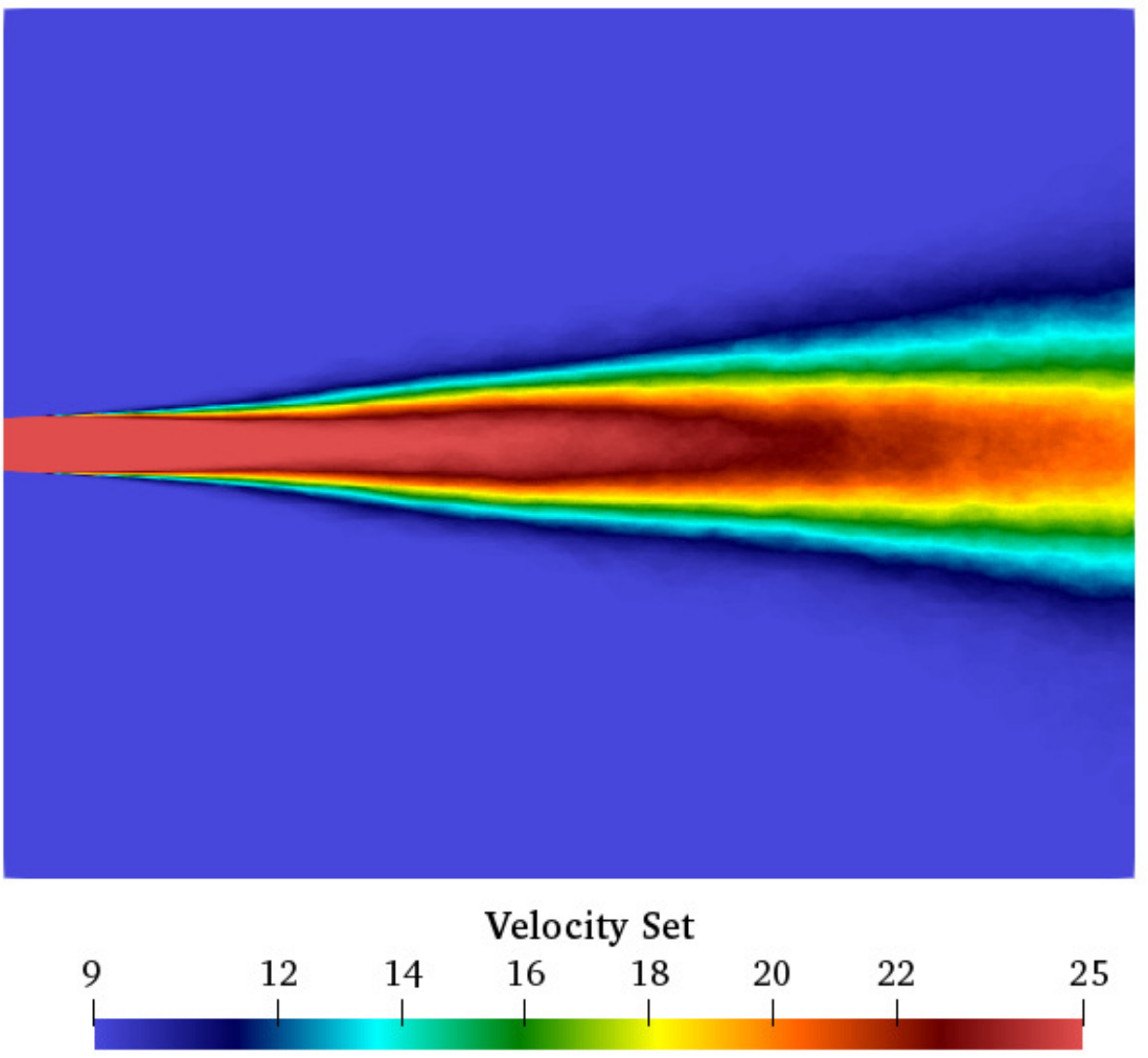} 
\includegraphics[width=0.45\textwidth]{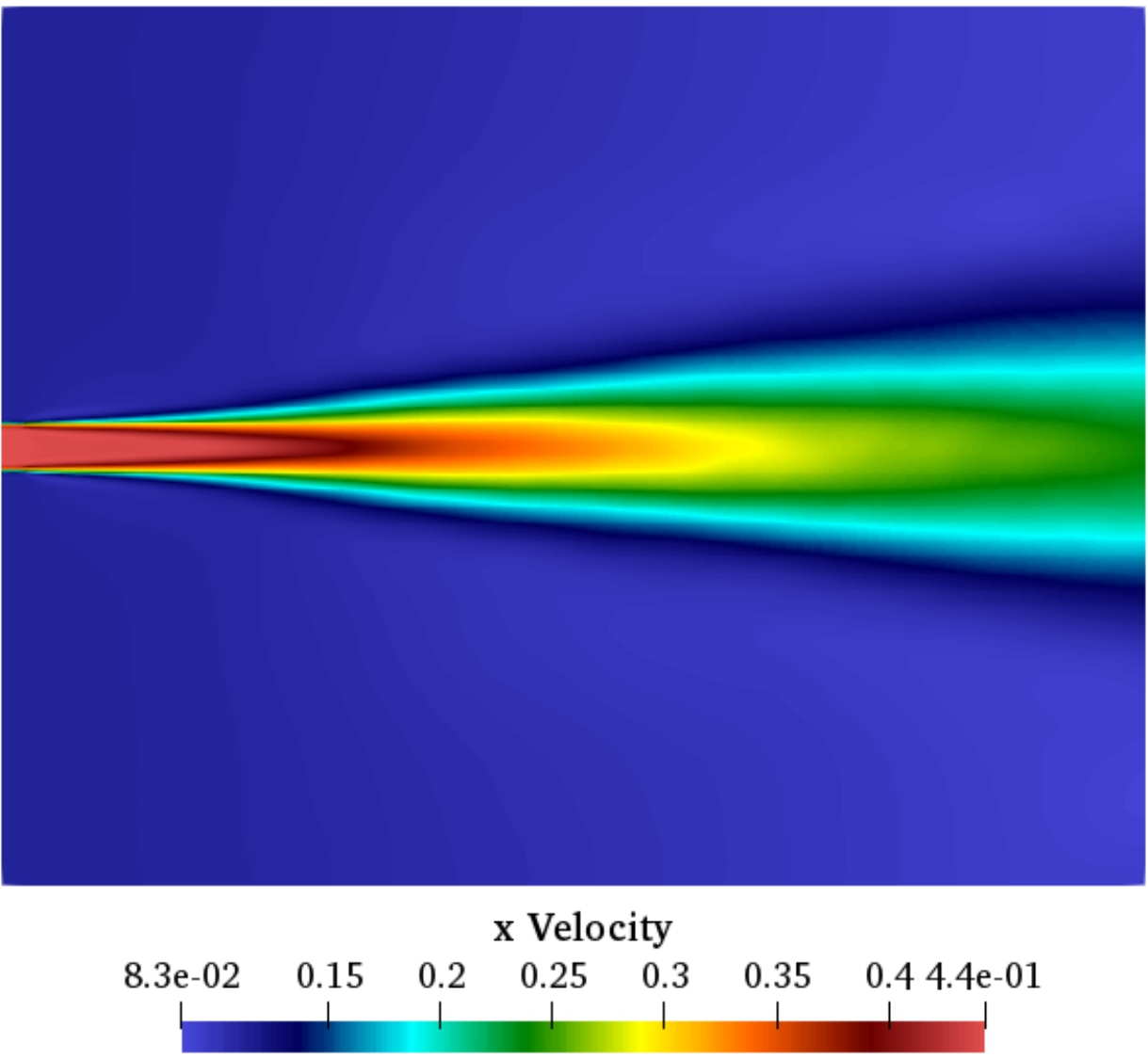}

\caption{Simulation of athermal jet flow with coupled D2Q9/D2Q25. (Top) Time average of the velocity set; (Bottom) Time average of the x-velocity.}
\label{fig:JetFlow_avg}
\end{figure}

The multi-scale model uses the D2Q9 and D2Q25 velocity sets, at a fixed reference frame at rest. The refinement criterion is based on the local Mach number and set to $Ma_{\text{thr}}=0.3$. Figure \ref{fig:JetFlow_Inst} shows the instantaneous x-velocity field and the corresponding spatial distribution of the velocity set at the same time step. The time average results for the velocity set and the x-velocity are shown in figure \ref{fig:JetFlow_avg}, where it is evident that the D2Q25 is dominant across the center-line and near the jet inlet whereas the D2Q9 is used at low-speed regions of the flow. Note that the average ratio of the nodes using the high order velocity set is roughly 15\% and the discussion regarding the computational speed-up is presented in section \ref{ComputationalEff} .

\begin{figure}

\includegraphics[width=0.45\textwidth]{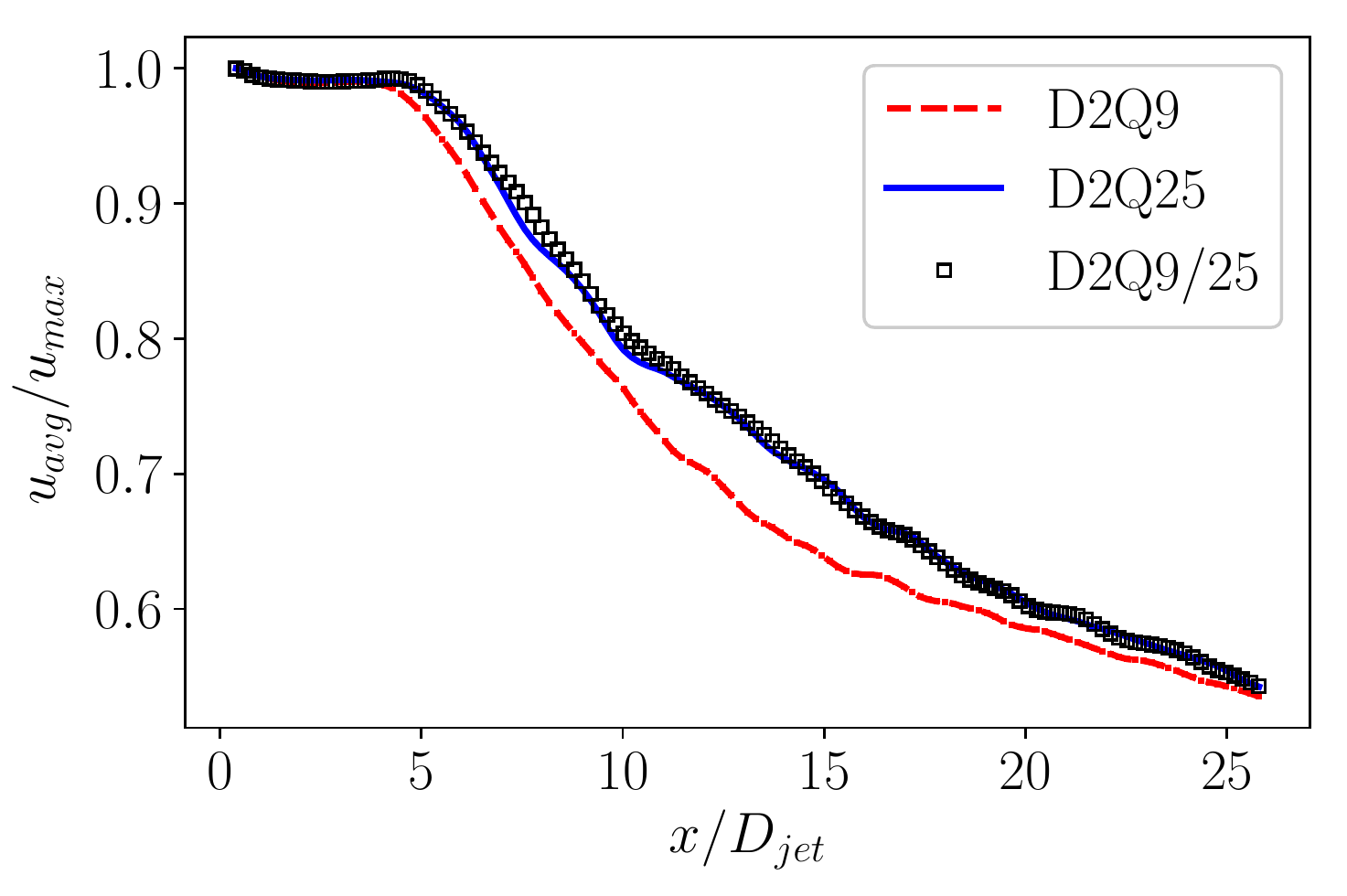} 
\includegraphics[width=0.45\textwidth]{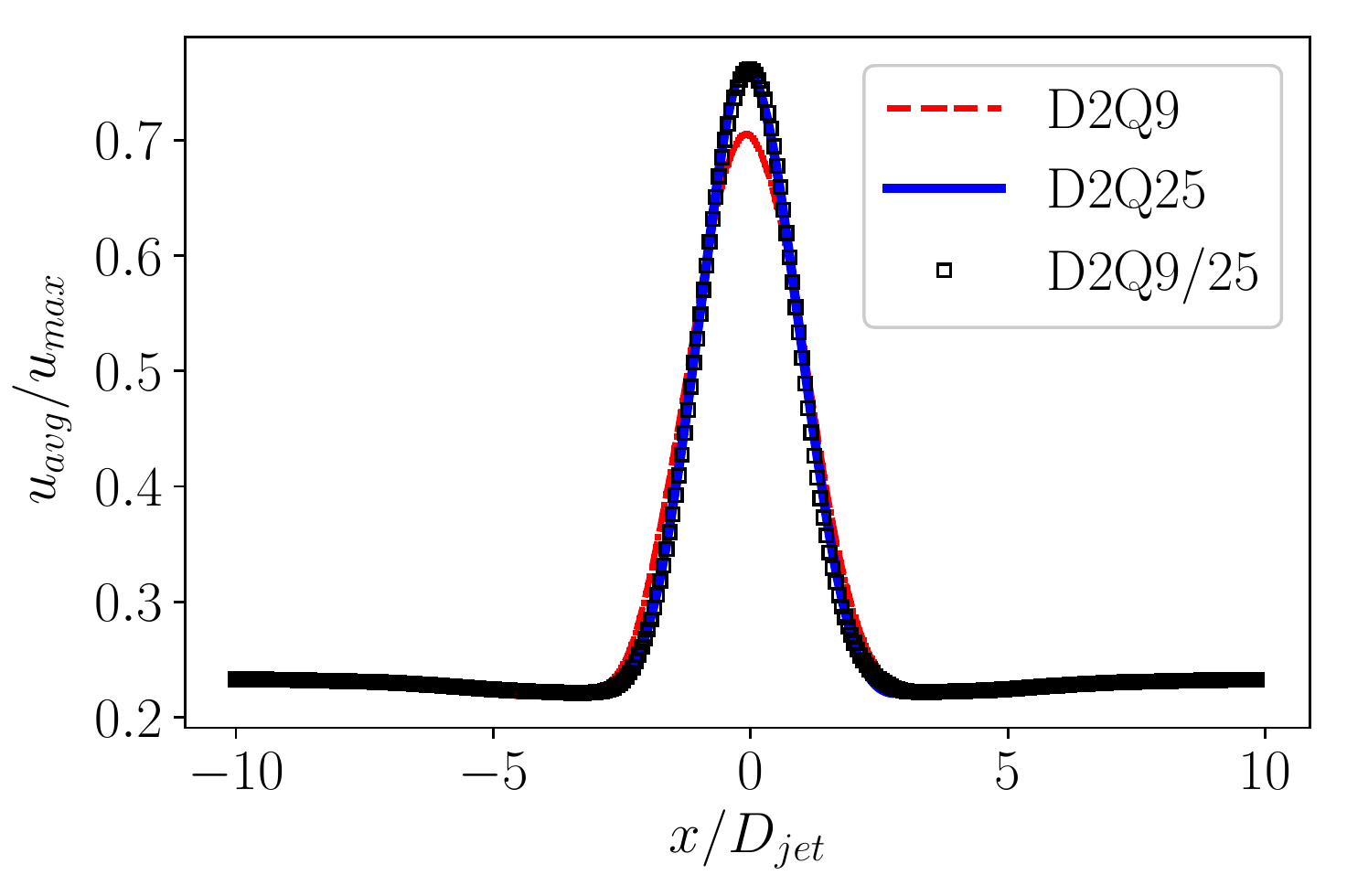}
\includegraphics[width=0.45\textwidth]{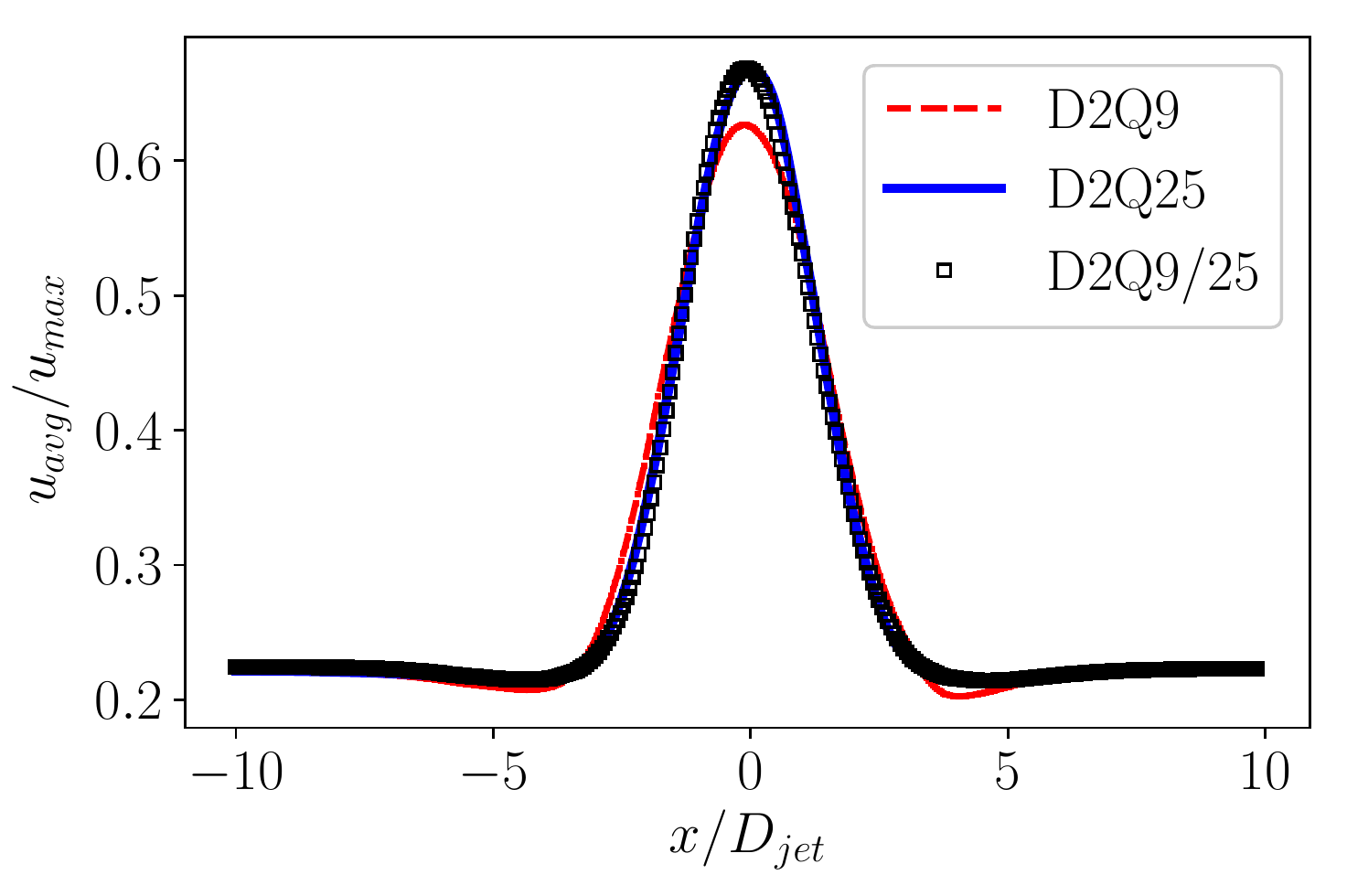}

\caption{Comparison of athermal jet flow with D2Q9, D2Q25 and coupled D2Q9/D2Q25. (Top) Mean streamwise velocity profiles along the centerline; (Middle) Mean streamwise velocity profiles at $x/D_{\text{jet}}= 12$; (Bottom) Mean streamwise velocity profiles at $x/D_{\text{jet}}= 16$  }
\label{fig:JetFlow_avgComp}
\end{figure}

In figure \ref{fig:JetFlow_avgComp}, the multi-scale simulation using the hybrid D2Q9/25 lattice is further compared with simulations in which the D2Q9 and D2Q25 were applied uniformly in the entire domain. The top plot shows the average x-velocity (normalised with the maximum velocity of the jet) across the center-line, while the middle and the bottom plots show the average x-velocity across planes normal to the flow, with distances from the inlet $x_{c}/D_{\text{jet}}=12,16$. It is apparent that the multi-scale  simulation is in good agreement with the simulations obtained with the D2Q25. In contrast the D2Q9 shows significant deviations with respect to D2Q25. This further validates the proposed multi-scale scheme.

\subsubsection{Shock structure}

The shock structure problem is a classical problem in kinetic theory of gases, in which non-equilibrium phenomena dominate the flow. It is well known that since the thermodynamic variables vary on a scale of few mean free paths, traditional continuum equation such as Navier-Stokes-Fourier equations fail to correctly describe the shock wave structure. 

The numerical setup consists a quasi one-dimensional plane shock wave, with an initial step at the center of the computational domain. The underlying model consists of the two-population realisation of the PonD method, with Prandtl number fixed to $Pr=2/3$ and adiabatic exponent of monoatomic ideal gas $\gamma=5/3$. The upsteam and downstream flow values are connected through the Rankine-Hugoniot conditions \cite{Anderson}. The upstream mean free path for hard sphere molecules is defined as,
\begin{equation}
\lambda_1=\frac{16}{5\sqrt{2\pi\gamma}} \left[ \frac{\mu_1\alpha_1}{p_1} \right],
\end{equation}
where $p_1,\alpha_1,\mu_1$ are the pressure, speed of sound and the viscosity of the gas upstream of the shock respectively. The viscosity varies with the temperature as,
\begin{equation}
    \mu=\mu_1\left(\frac{T}{T_1}\right)^s,
\end{equation}
where for the case of hard spheres $s=0.5$.

The steady-state non-dimensional density, temperature, normal stress and heat flux are defined as follows,
\begin{equation}
\rho_n=\frac{\rho-\rho_1}{\rho_2-\rho_1}, T_n=\frac{T-T_1}{T_2-T_1}, \hat{\sigma}_{xx}=\frac{\sigma_{xx}}{p_1}, \hat{q}_x=\frac{q_x}{p_1\sqrt{2T_1}},
\end{equation}
where the subscripts 1 and 2 indicate the upstream and downstream values w.r.t. the shock wave.

\begin{figure}

\includegraphics[width=1\linewidth]{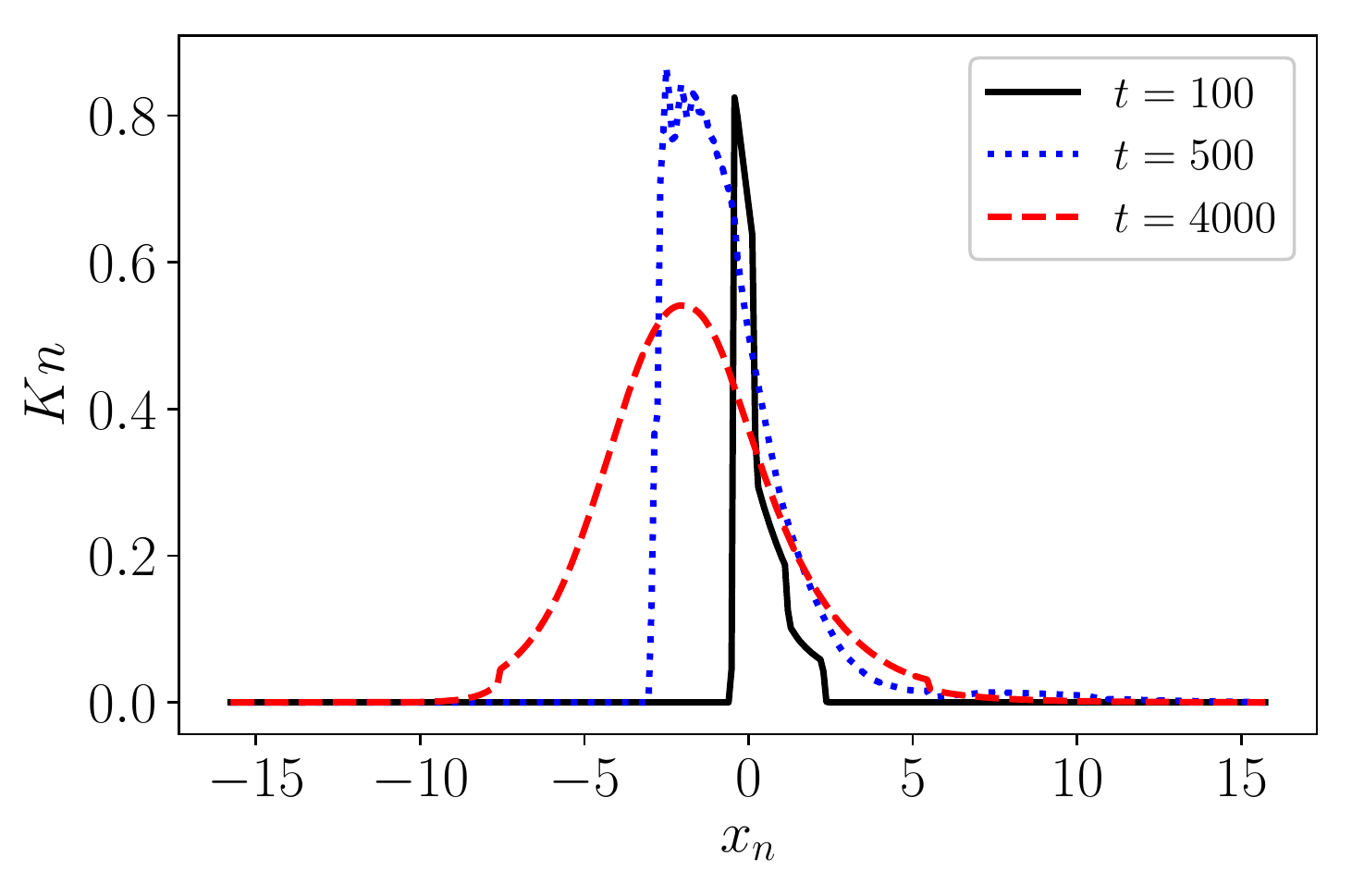}

\includegraphics[width=1\linewidth]{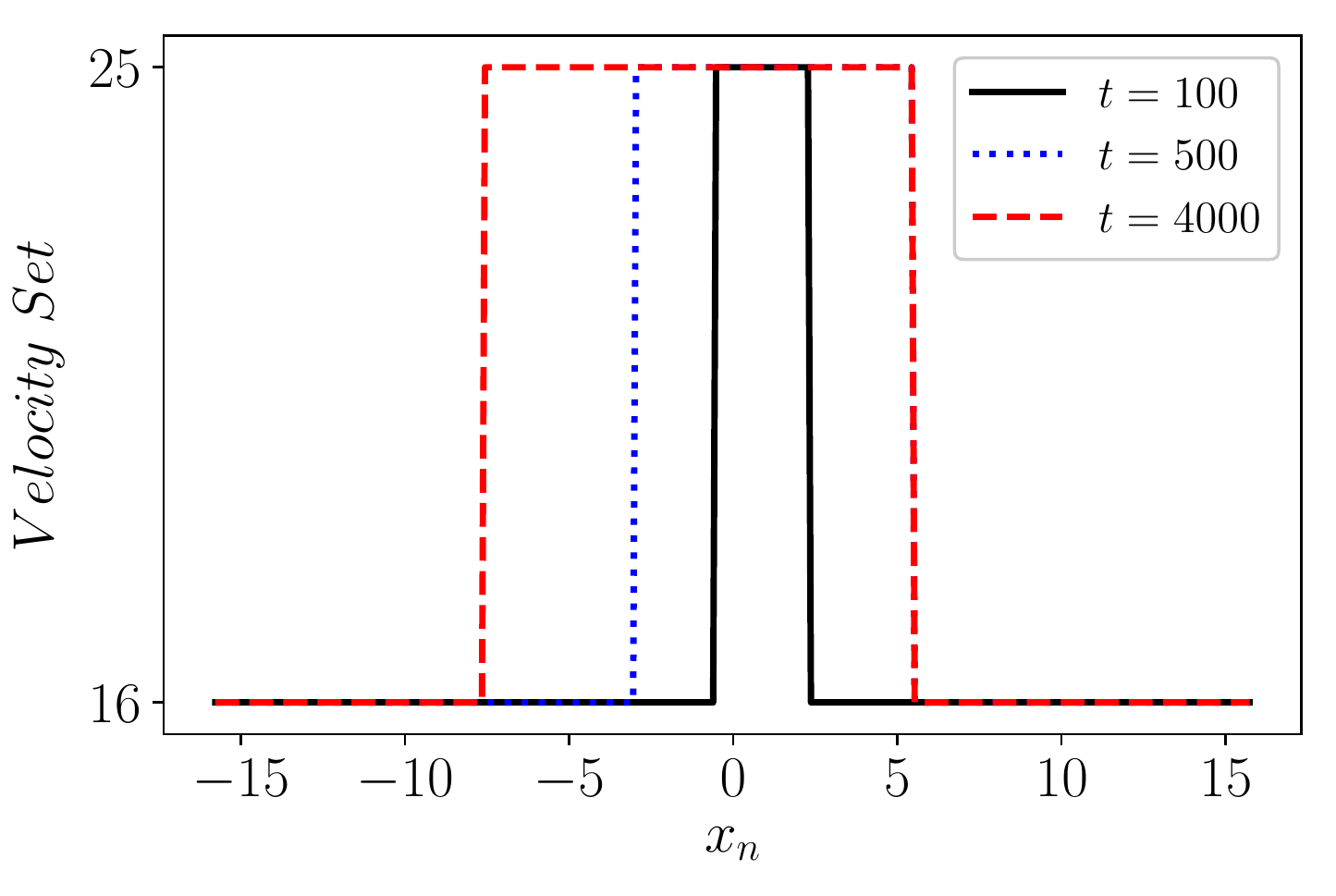}

\caption{Evolution of Knudsen profile (top) and velocity set (bottom) at times $t=100,500,4000$.}
\label{fig:ShockEvol}

\end{figure}

The D2Q16 and D2Q25 velocity sets are coupled for the computation of the shock structure. The high order velocity is used inside the region of the shock wave and the low order in the rest of the computational domain. The refinement criterion is based on the local Knudsen number, which can be computed as follows,
\begin{equation}
    Kn= \frac{\lambda}{L}, L=\left|\frac{\phi}{d\phi /dx}\right|,
\end{equation}
where $\phi$ is density (in general other macroscopic fields can be used such as Mach number). The threshold value was set to $Kn_{\text{thr}}=0.02$ , a typical value suggested by the literature \cite{DSCM_BirdBook,multiscale}. The simulations were carried out with a quasi one-dimensional setup with the resolution of the lattice $\Delta x$ corresponding to $0.02\lambda_1$. The velocity set in the domain is initialized with the D2Q16 lattice and wherever the local Knudsen number exceeds the threshold value is refined to D2Q25. The time evolution of the velocity set and the computed Knudsen number is shown in figure \ref{fig:ShockEvol}. The D2Q25 is initially concentrated in the middle of the domain and gradually expands towards the entire shock transition layer.

\begin{figure}

\includegraphics[width=1\linewidth]{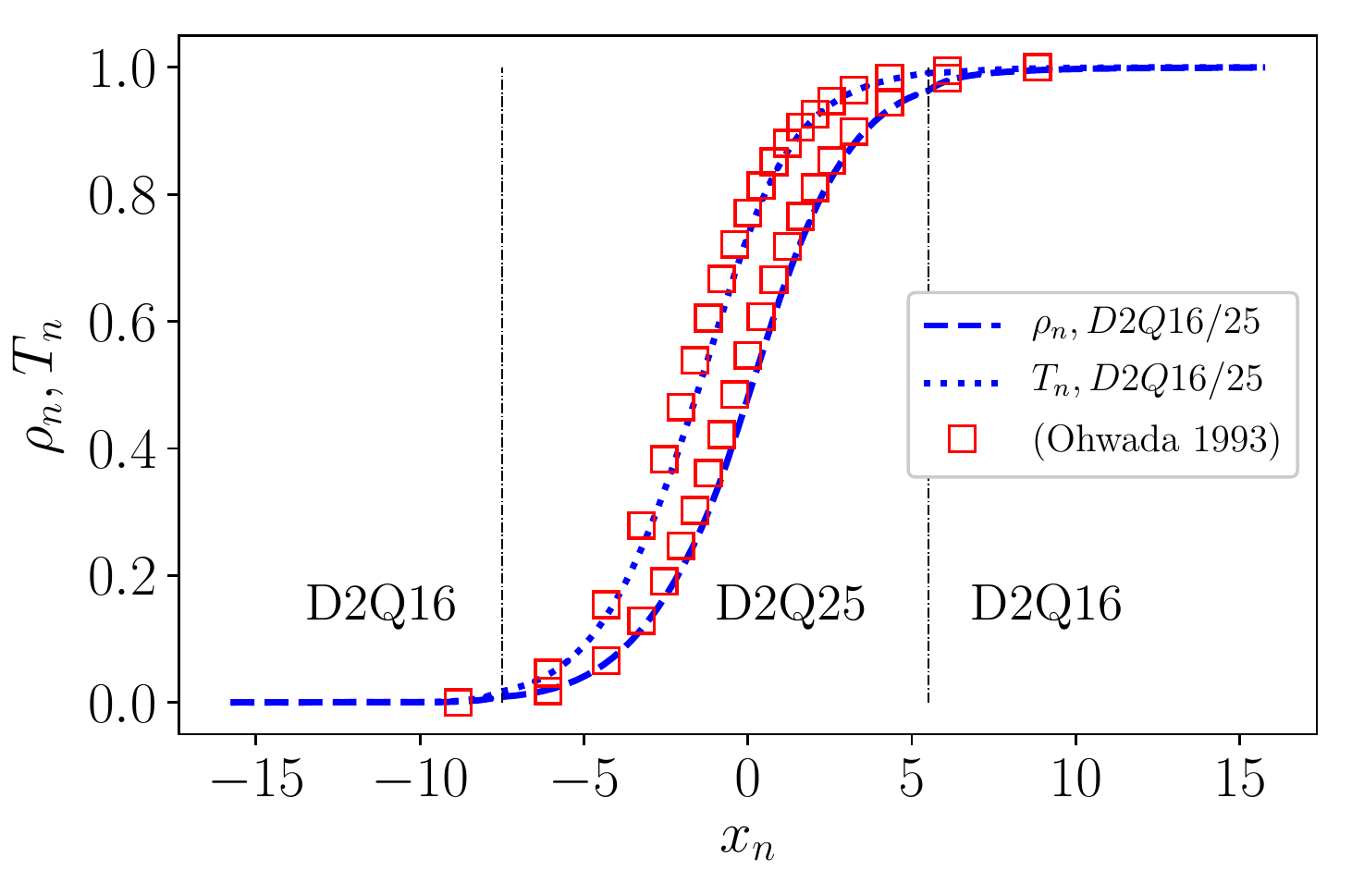}

\includegraphics[width=1\linewidth]{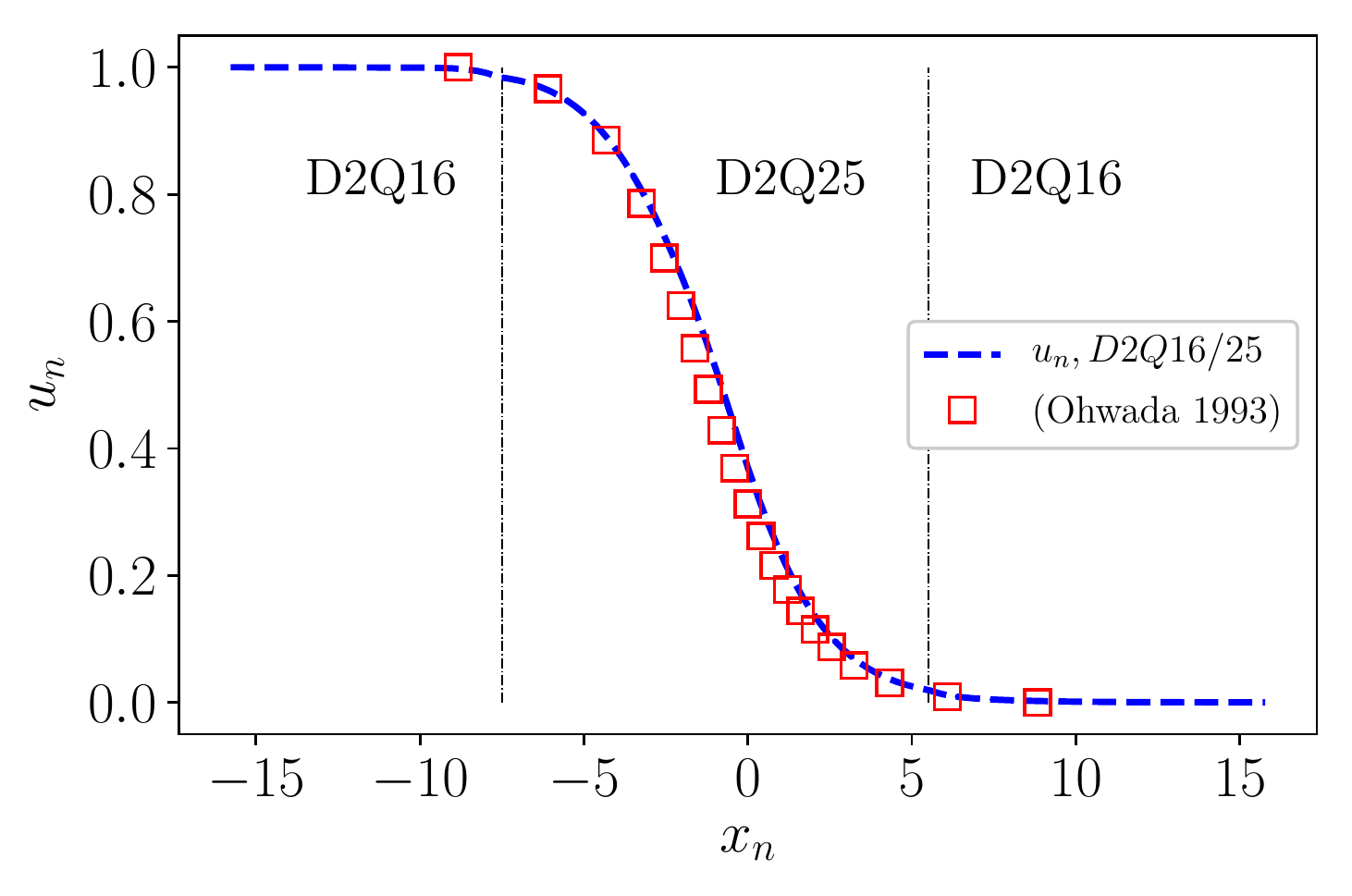}

\includegraphics[width=1\linewidth]{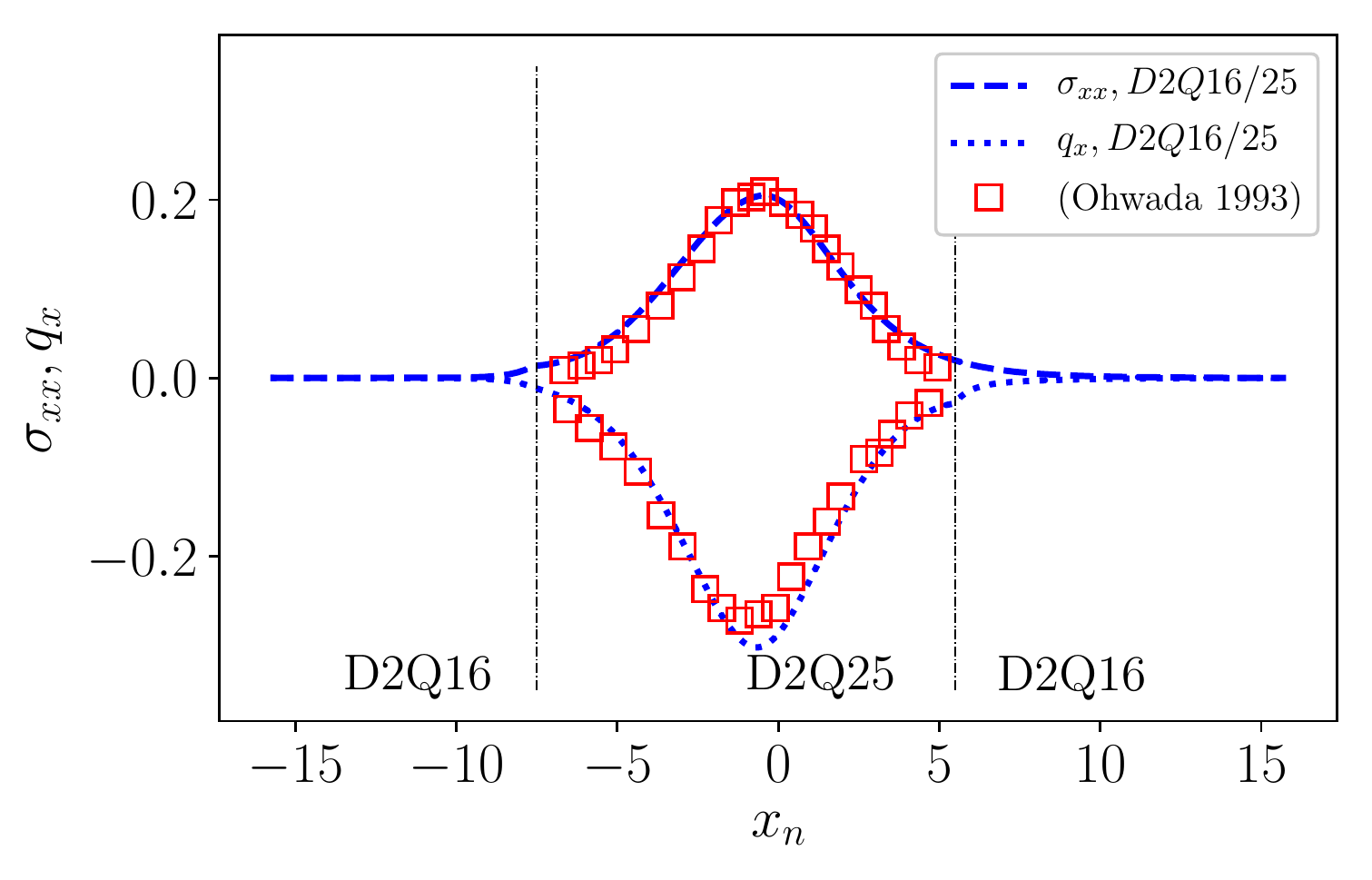}

\caption{(Multi-scale D2Q16/25) Density,temperature, velocity, normal stress and heat flux profiles for $Ma=1.6$, compared with results of Ohwada \cite{ohwada}. The vertical dotted lines on the plot indicate the high-order lattice.}
\label{fig:Shock1}

\end{figure}

The steady-state results for $Ma=1.6$  are shown in figure \ref{fig:Shock1} and compared with the results of Ohwada \cite{ohwada}. The origin of the coordinate system is the point with $\rho_n=0.5$ and x is non-dimensionalized with $0.5\sqrt{\pi}\lambda_1$. It can be seen that the density, temperature and velocity profiles match very well with the reference data. The normal stress and heat flux profiles are also in good agreement with the reference data.

\subsection{Micro-Couette flow}

The simulation of microscale flows is of both practical and theoretical interest especially when the Knudsen numbers is not negligible. A classical case to test the behaviour of numerical solvers for this regime is the shear driven Couette flow. It has been established that the choice of the discrete velocities set as well as the boundary conditions are of paramount importance for the accurate description of the fluid near the walls. In particular the standard D2Q9 lattice does not suffice to capture the Knudsen layer and significant deviations occur for the velocity profile at finite Knudsen numbers. Furthermore it has been reported that even-orders velocity sets perform significantly better than odd-orders \cite{SlipPlateMeng}.

\begin{figure}[h]
    \centering
   \includegraphics[width=0.45\textwidth]{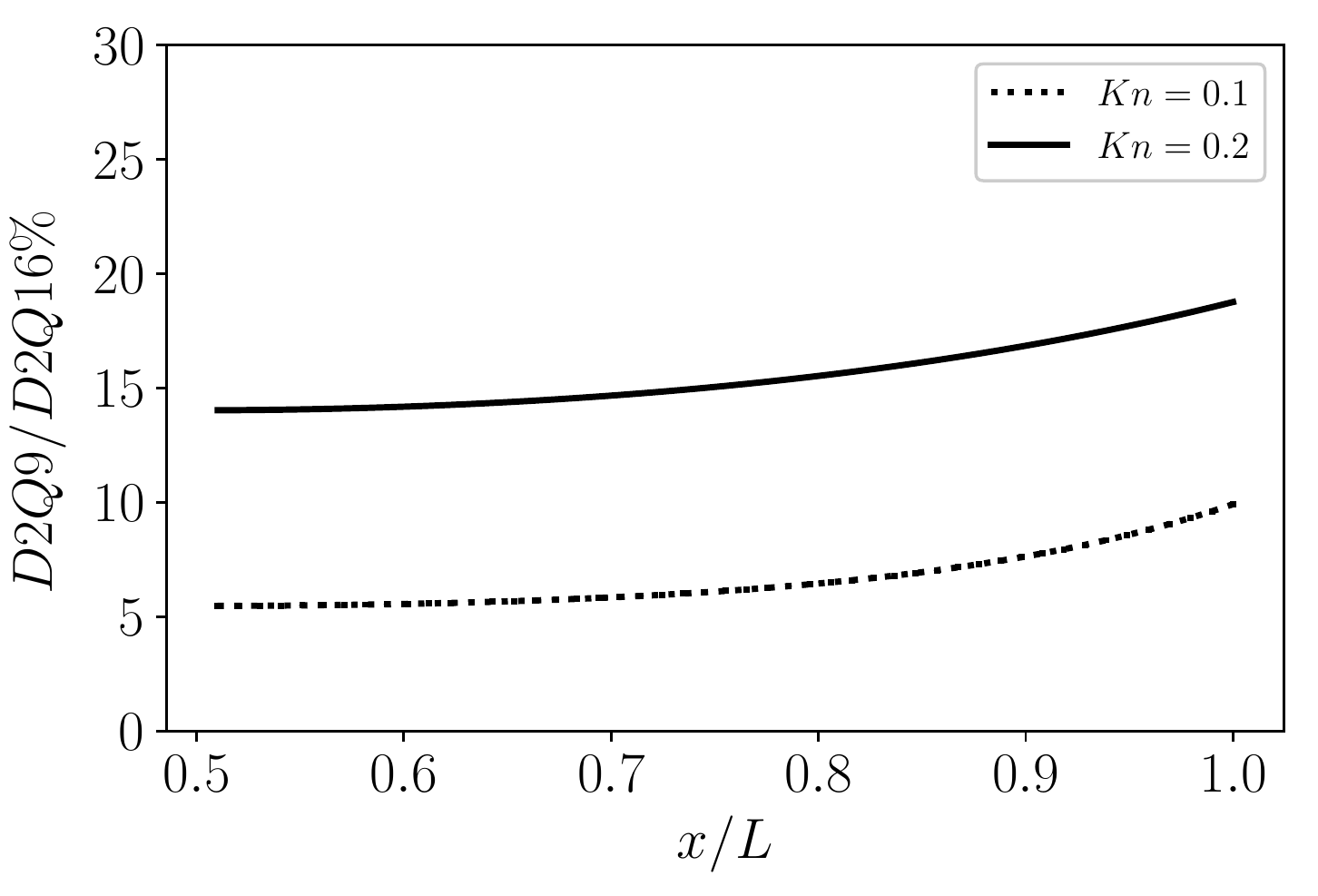}
   
    \caption{Relative error between D2Q9 and D2Q16 for shear driven Couette flow, based on analytical solution \cite{AnsumaliPlateSlip}.}
    \label{fig:SlipPlateAnalyticalError}
\end{figure}

At this point, we must stress that the micro-Couette flow is not a true multi-scale problem. Indeed the Knudsen number is uniform in the entire domain and 
in contrast to shock-structure problem, where the Knudsen number varies within the shock transition layer. For a quantitative discussion, we use the exact solution of the micro-Couette flow, which was been found analytically for D2Q9 and D2Q16 lattices in \cite{AnsumaliPlateSlip}. 
The authors concluded that while the D2Q9 can predict a slip-flow solution, it fails in the transient regime ($Kn\gtrapprox 0.1$). 
The D2Q16 lattice on the other hand improves the accuracy of the solution considerably and predicts the boundary Knudsen layer in qualitative agreement with kinetic theory \cite{CercignBook}. Figure \ref{fig:SlipPlateAnalyticalError} shows the relative error of the analytical solutions obtained by D2Q9 and D2Q16 for Knudsen numbers 0.1 and 0.2. At a constant Knudsen number the relative error has a constant value in the main flow and grows in the vicinity of the wall boundaries. Increasing the Knudsen number leads to magnification of the error in the main flow as well as extending the influence of the wall due to the boundary Knudsen layer.

The setup of the Couette test case consists of two plates surrounding the fluid, which move with opposite velocities $u_w=\pm0.1$ and same temperature $T_w=T_L$. The simulations were performed with a quasi-one dimensional $[300 \times 4]$ grid. We test the multi-scale scheme with the D2Q9 velocity set in the main flow and the D2Q16 near the wall boundaries. The underlying model consists of the two-population realisation of the PonD method, with Prandtl number fixed to $Pr=2/3$ and adiabatic exponent of monoatomic ideal gas $\gamma=5/3$. Diffusive boundary conditions are implemented to efficiently capture the gas-wall interactions \cite{AnsumaliKBC,SlipPlateMeng}.

\begin{figure}[h]

\includegraphics[width=0.45\textwidth]{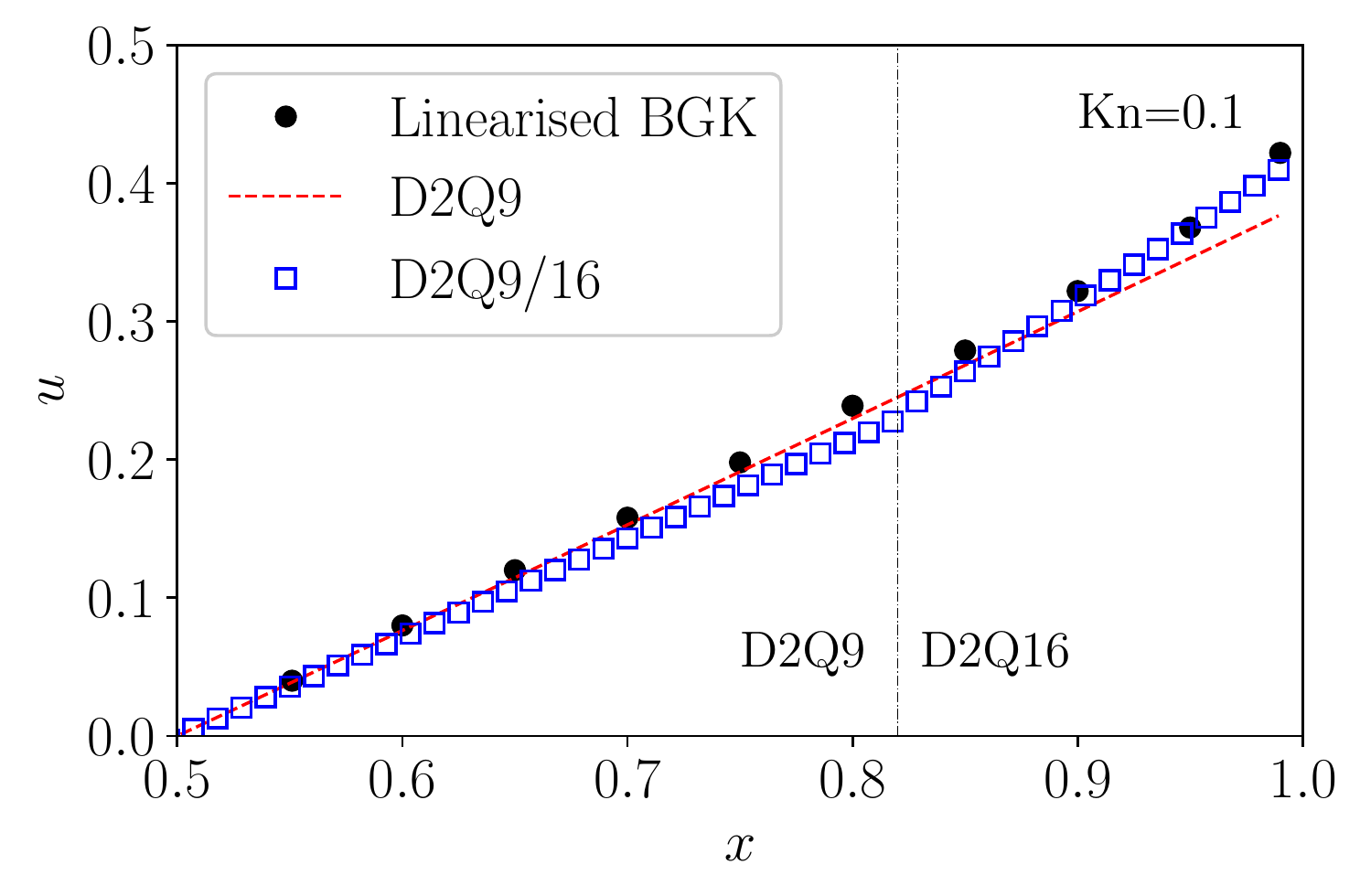}
\includegraphics[width=0.45\textwidth]{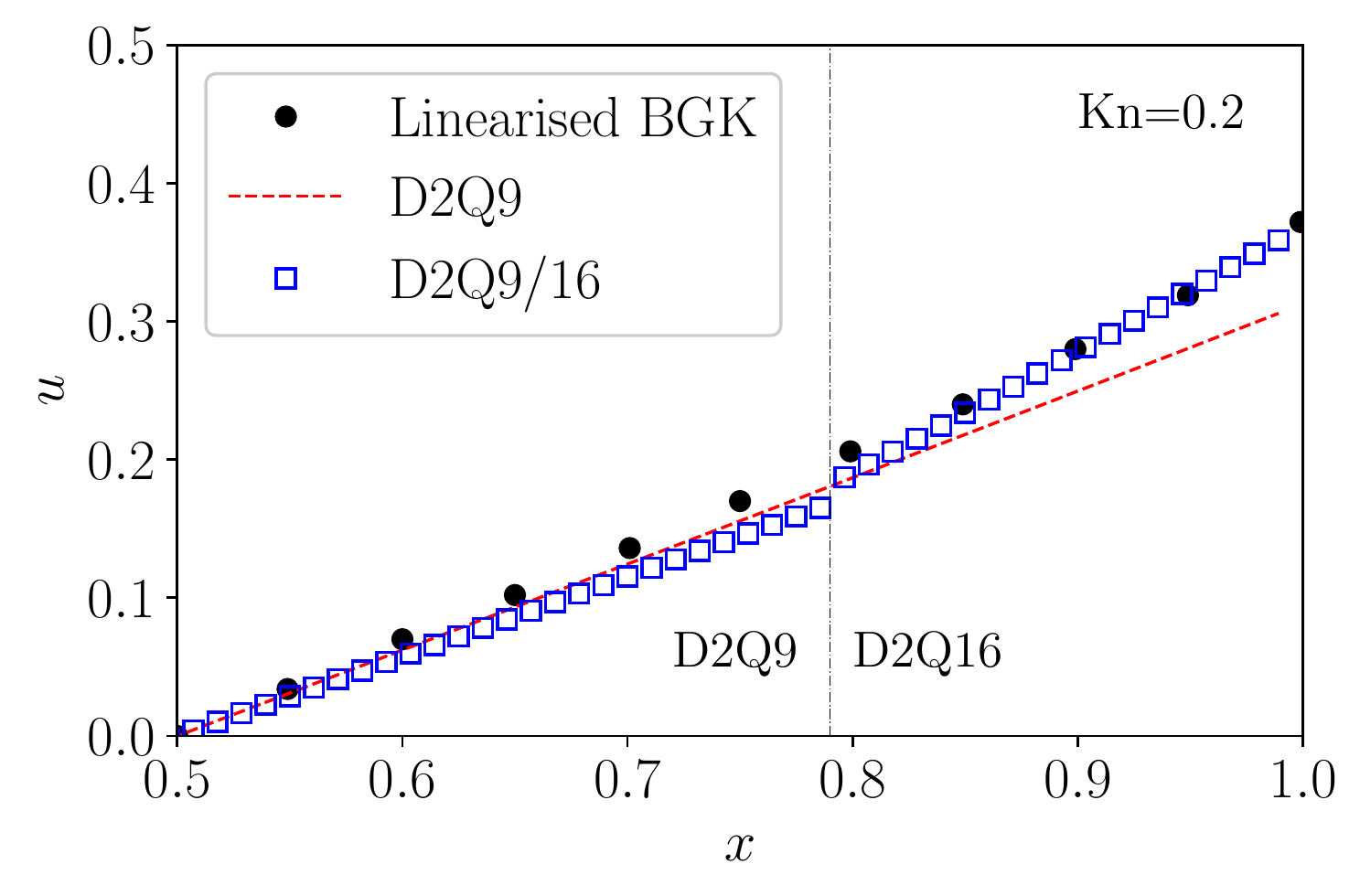}
\includegraphics[width=0.45\textwidth]{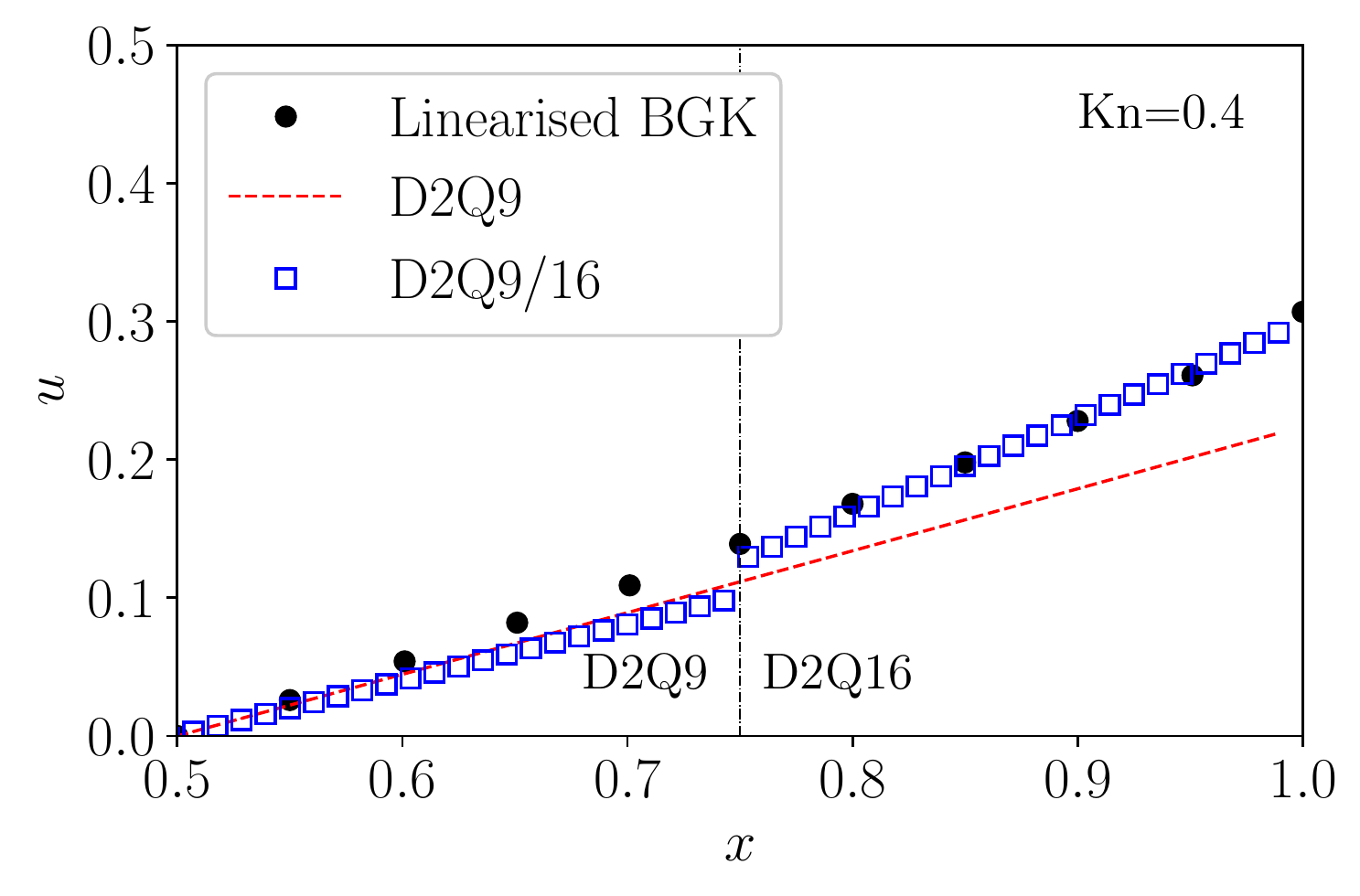}

\caption{Shear driven Couette flow with D2Q9, D2Q16 and coupled D2Q9/16. Profiles of the normalised velocity for Knudsen numbers $Kn=0.1$ (top); $Kn=0.2$ (middle); $Kn=0.2$ (bottom)}
\label{fig:SlipPlate}
\end{figure}

 The results for the non dimensional velocity (normalised with the difference of the wall velocities) for Knudsen numbers equal to 0.1, 0.2 and 0.4 are presented in figure \ref{fig:SlipPlate}, in which comparison is made with results from linearised BGK \cite{SlipPlateMeng}. The dotted vertical lines on the plots indicate the D2Q9/D2Q16 interface for the multi-scale simulations. The position of the D2Q9/D2Q16 interface is set according to the theoretical error analysis (1\% above the main flow error).
 
We observe that for all Knudsen numbers the multi-scale scheme matches well with D2Q9 solution in the main flow and also with the reference results near the wall boundaries, due to the local use of the D2Q16 lattice. However, a small discontinuity develops near the interface region of the different velocity sets, which becomes more prominent with increasing Knudsen number. This is an expected behaviour, caused by the discrepancy between the solutions of D2Q9 and D2Q16, as shown in Fig. \ref{fig:SlipPlateAnalyticalError}.

\subsection{Computational efficiency}
\label{ComputationalEff}

\begin{figure}

\includegraphics[width=0.5\textwidth]{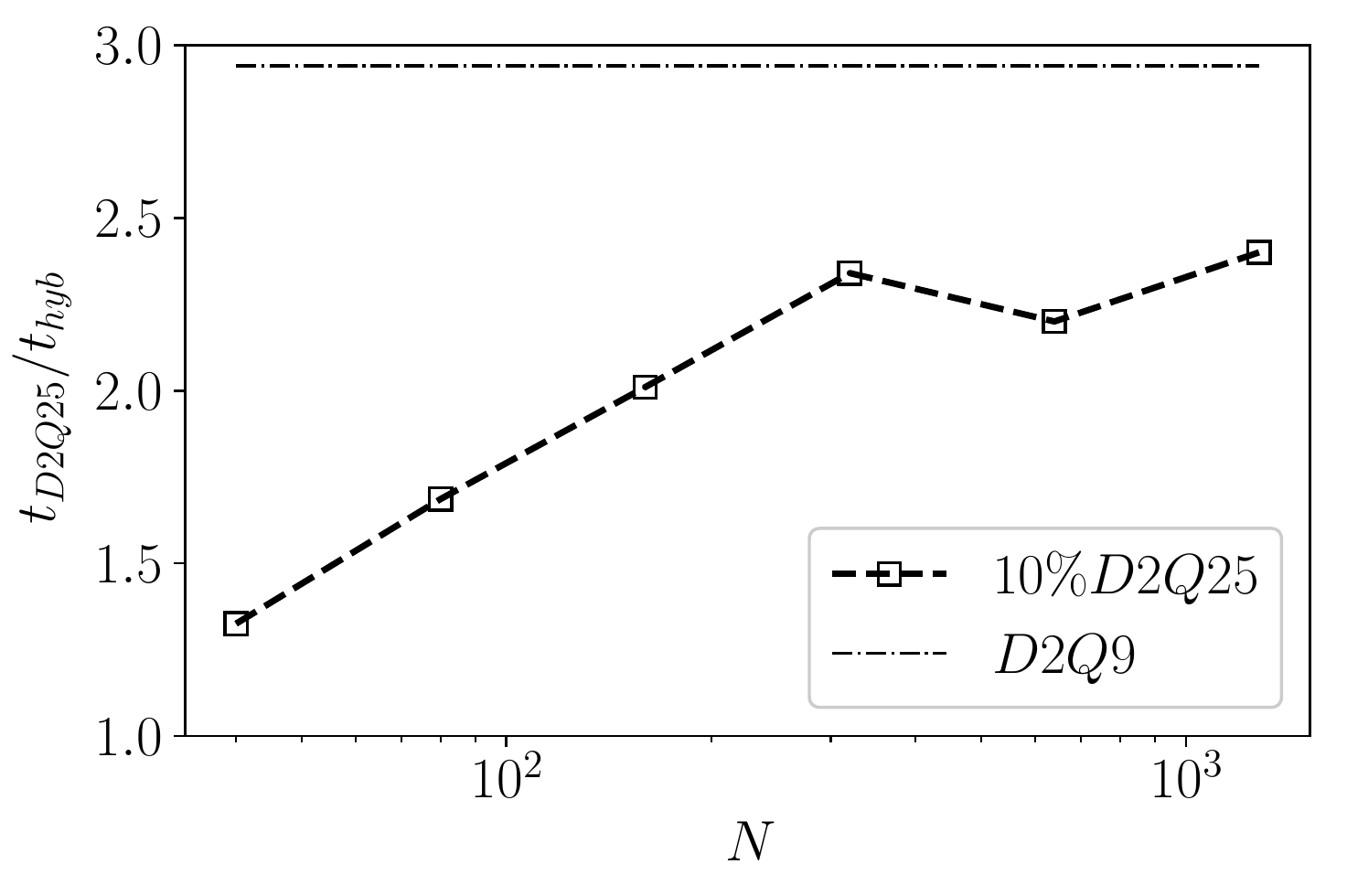}

\caption{Speed up ratio $t_{D2Q25}/t_{hyb}$ as a function of the domain size N. Ratio of high-order nodes (D2Q25) is fixed to $10 \%$. }
\label{fig:SpeedUp}

\end{figure}

In this section, we investigate the increase of computational efficiency when using the multi-scale scheme.  
To that end, a computational domain of size $(N \times N)$ is decomposed in two rectangular zones which use the high-order D2Q25 and low-order D2Q9 velocity sets respectively. The ratio of the CPU times $t_{\text{D2Q25}}/t_{\text{hyb}}$, where $t_{\text{D2Q25}}$ refers to the time of a globally used D2Q25 and $t_{\text{hyb}}$ to the hybrid scheme, is measured as a function of the domain size $N$, keeping the ratio of the high-order nodes to $10 \%$ of the computational domain fixed. The simulations were carried out on a eight-core PC (Intel Core i7-9700@3GHz) and the timings are shown in Figure \ref{fig:SpeedUp}.
The speedup that can be observed for the simulation using the D2Q9 lattice in the entire domain is $t_{D2Q25}/t_{D2Q9}=2.9$. Thus, the maximum speedup that can be achieved when the ratio of D2Q25 is fixed to $10\%$ is $(t_{\text{D2Q25}}/t_{\text{hyb}})_{\text{max}}=2.43$. For small domain sizes the computational cost due to the lifting and projection transformations at the interface nodes limits the speedup but for larger domains ($N > 100$) significant speedup is observed, saturating close to the optimal value at around 2.4. The proposed multi-scale scheme can therefore lead to significant savings in CPU time.

\section{Conclusion}
\label{Conclusion_section}

In this work, we presented a novel multi-scale scheme with an adaptive velocity set, according to a refinement criterion based on the local Mach or Knudsen number. Velocity sets of different order are coupled through the lifting and projection operators. Both operators involve only local computations, which results in a robust and flexible adaptive velocity refinement. The multi-scale scheme can be implemented with either a static or co-moving reference frame and with different models (single/double population, quasi-equilibrium models). The numerical results with a variety of flows validated the accuracy and efficiency of the proposed scheme. 
This work focused on 2D applications, but the lifting and projection, underlying the coupling scheme, hold also in three dimensions. 
Thorough investigation of the proposed scheme in three dimensions is left for future work.

\begin{acknowledgments}
This work was supported by European Research Council (ERC) Advanced Grant No. 834763-PonD. 
Computational resources at the Swiss National  Super  Computing  Center  CSCS  were  provided  under grant No. s897.
\end{acknowledgments}

\appendix

\section{Convergence order}

\begin{figure}

\includegraphics[width=0.5\textwidth]{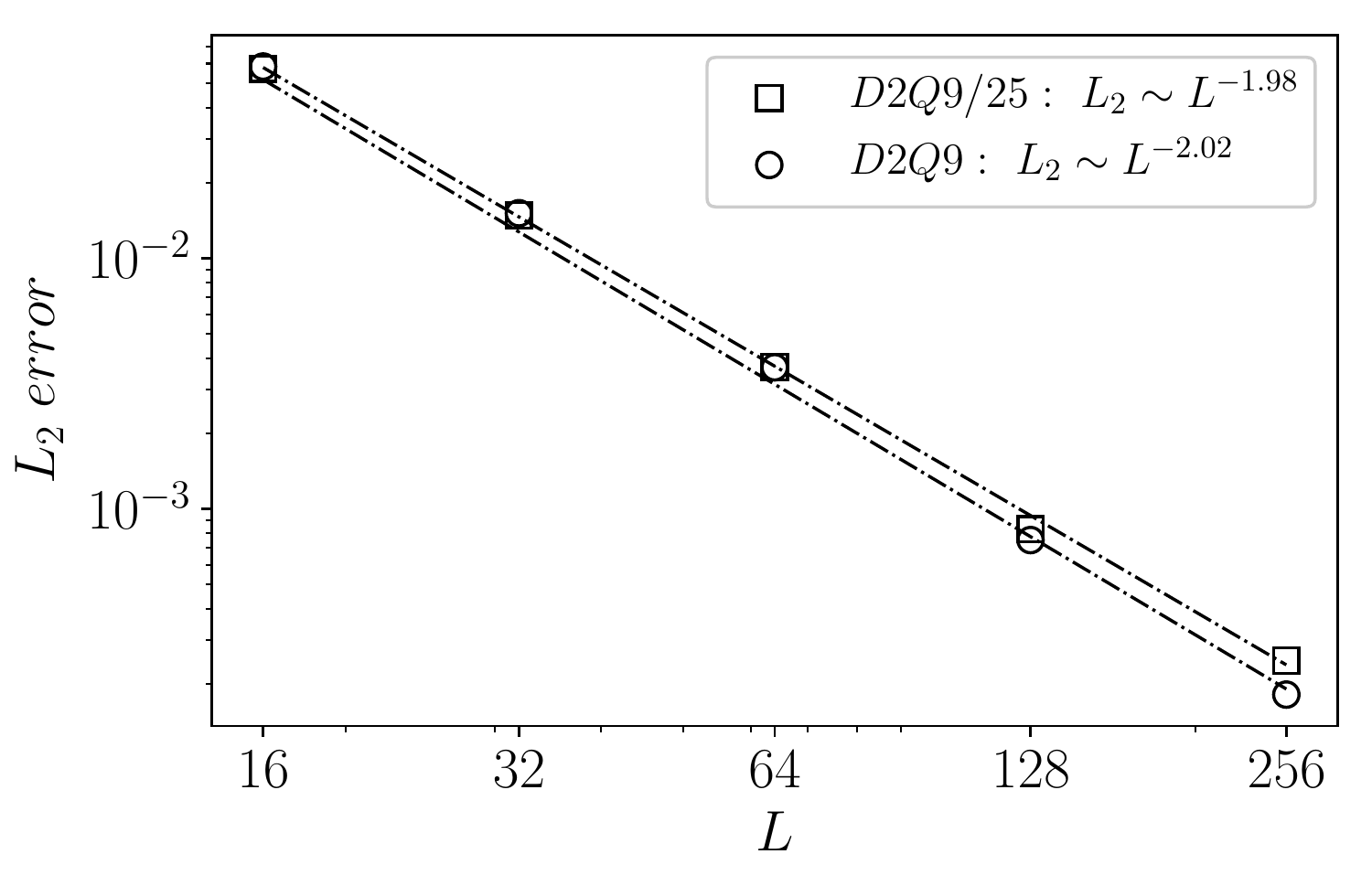}

\caption{Scaling of $L_2$ error for D2Q9 and coupled D2Q9/25 for Taylor-Green vortex.}
\label{fig:errorConv}

\end{figure}

The accuracy in space of the multi-scale scheme is studied with the incompressible Taylor-Green vortex. The high-order D2Q25 is used inside a circular disk centered in the domain and the D2Q9 is used outside. Figure \ref{fig:errorConv} shows the scaling of the $L_2$ error, which indicates that the multi-scale scheme retains second-order accuracy.

\section{PonD transformation matrix}

The implementation of the PonD method relies on the transformation of a population vector in different reference frames, $\lambda=\{ \bm{u},T\}$ and $\lambda'=\{ \bm{u}',T'\}$. The transformation is a linear operation, based on moment matching,

\begin{equation}
    f^{\lambda'}=\mathcal{M}_{\lambda'}^{-1}\mathcal{M}_{\lambda}f^\lambda=\mathcal{G}_{\lambda}^{\lambda'}f^\lambda.
\end{equation}

Below we provide a simple script to obtain analytical expression for the transformation matrix for the case of D2Q9 velocity set, using the symbolic toolbox of Matlab,

\begin{widetext}

\begin{verbatim}
    syms ux uy th; %
    syms ux_p uy_p th_p; %
    
    syms M M_p; %
    
    Q=9;
    cx=[0 1 0 -1 0 1 -1 -1 1];
    cy=[0 0 1 0 -1 1 1 -1 -1];
    
    pow_x=[0,1,0,1,2,0,2,1,2];
    pow_y=[0,0,1,1,0,2,1,2,2];
    
    for i=1:1:Q
        for j=1:1:Q
            M(i,j)=(th*cx(j)+ux)^(pow_x(i))*(th*cy(j)+uy)^(pow_y(i));
            M_p(i,j)=(th_p*cx(j)+ux_p)^(pow_x(i))*(th_p*cy(j)+uy_p)^(pow_y(i));
        end
    end
    
    
    TransfMatrix=inv(M_p)*M;
    
    
\end{verbatim}

\end{widetext}

The same script with can be applied for other velocity sets, with a redefinition of the lattice velocities and the list of moments.

\section{Lifting and projection operations}

To illustrate the coupling scheme we consider the lifting and projection operations, for the case of D2Q9-D2Q16 coupling. The velocity set of the D2Q9 $\{\bm{v}_i^9 \}$ lattice is formed from the tensor product of the discrete velocities in 1D $(-c,0,c)$ and the D2Q16 lattice  $\{\bm{v}_i^{16} \}$ from $(-c_2,-c_1,c_1,c_2)$.

For the lifting operation, we need to construct the higher-order moment vector $M_{9 \rightarrow 16}=[M_{00},M_{10},\cdots,M_{33}]$, given the lower-order population vector $f$. The 9 first moments are directly computed from $f$,

\begin{equation}
    M_{kl}=\sum_{i=0}^{i=8}{(v^9_{ix})^k (v^9_{iy})^l f_i},
\end{equation}

where $k,l \in [0,2]$. The remaining moments are computed from the higher-order equilibrium population $f^{eq,16}$,
\begin{equation}
    M_{kl}=\sum_{i=0}^{i=15}{(v^{16}_{ix})^k (v^{16}_{iy})^l f^{eq,16}_i}.
\end{equation}

 Finally, the inverse of the moment matrix $\mathcal{M}^{-1}_{16}$ is applied to $M^{9 \rightarrow 16}$,
\begin{equation}
    f_{9 \rightarrow 16}=\mathcal{M}^{-1}_{16} M_{9 \rightarrow 16}.
\end{equation}

For the projection operation we compute the moment vector $M_{16 \rightarrow 9}=[M_{00},M_{10},\cdots,M_{22}]$,given the higher-order population vector $f$ as,

\begin{equation}
    M_{kl}=\sum_{i=0}^{i=15}{(v^{16}_{ix})^k (v^{16}_{iy})^l f_i},
\end{equation}

where $k,l \in [0,2]$. The inverse of the moment matrix $\mathcal{M}^{-1}_{9}$ is applied to $M^{16 \rightarrow 9}$,
\begin{equation}
    f_{16 \rightarrow 9}=\mathcal{M}^{-1}_{9} M_{16 \rightarrow 9}.
\end{equation}

The analytical forms of the matrices $\mathcal{M}^{-1}_{9},\mathcal{M}^{-1}_{16}$ can be obtained from the following Matlab script,

\begin{widetext}

\begin{verbatim}

    syms M_min M_max;
    
    syms c;
    Q_min=9;
    cx_min=[0 c 0 -c 0 c -c -c c];
    cy_min=[0 0 c 0 -c c c -c -c];
    
    syms c1 c2;
    Q_max=16;
    cx_max=[c1 c1 -c1 -c1 c2 c2 c1 c1 -c1 -c1 -c2 -c2 c2 c2 -c2 -c2];
    cy_max=[c1 -c1 c1 -c1 c1 -c1 c2 -c2 c2 -c2 c1 -c1 c2 -c2 c2 -c2];    
    
    pow_x=[0,1,0,1,2,0,2,1,2,0,3,1,3,2,3,3];
    pow_y=[0,0,1,1,0,2,1,2,2,3,0,3,1,3,2,3];
    
    for i=1:1:Q_min
        for j=1:1:Q_min
            M_min(i,j)=(cx_min(j))^(pow_x(i))*(cy_min(j))^(pow_y(i));
        end
    end
    
    for i=1:1:Q_max
        for j=1:1:Q_max
            M_max(i,j)=(cx_max(j))^(pow_x(i))*(cy_max(j))^(pow_y(i));
        end
    end    
    
    
    M_min_inv=inv(M_min);
    M_max_inv=inv(M_max);
    
    
\end{verbatim}

\end{widetext}

\bibliographystyle{unsrt}

\providecommand{\noopsort}[1]{}\providecommand{\singleletter}[1]{#1}%

\end{document}